\documentclass[superscriptaddress,prb,amsmath,twocolumn,amssymb]{revtex4}
\usepackage{dcolumn}
\usepackage{amsfonts}
\usepackage{amssymb}
\usepackage{amsmath}
\usepackage{amsthm}
\usepackage{latexsym}
\usepackage{epsfig}
\usepackage{color}
\usepackage{lscape}
\usepackage{multirow}
\usepackage{braket}
\usepackage{bbm}
\usepackage{mathrsfs}
\usepackage{fancyhdr}
\usepackage{lastpage}
\usepackage{soul}
\usepackage{subfigure}
\usepackage[normalem]{ulem}
\usepackage[hidelinks=true]{hyperref}
\newcommand{\nohl}{\color{black}}

\newcommand{\be}{\begin{equation}}
\newcommand{\ee}{\end{equation}}
\newcommand{\ba}{\begin{array}}
\newcommand{\ea}{\end{array}}
\newcommand{\bea}{\begin{eqnarray}}
\newcommand{\eea}{\end{eqnarray}}

\begin{document}

\title{{
Large-scale silicon quantum photonics implementing arbitrary two-qubit processing
}}
\author{Xiaogang Qiang}
\affiliation{Quantum Engineering Technology Labs, H. H. Wills Physics Laboratory and Department of Electrical \& Electronic Engineering, University of Bristol, BS8 1FD,UK.}
\affiliation{State Key Laboratory of High Performance Computing, NUDT, Changsha 410073, China}
\affiliation{National Innovation Institute of Defense Technology, AMS, Beijing 100071, China.}
\author{Xiaoqi Zhou}
\email{zhouxq8@mail.sysu.edu.cn}
\affiliation{State Key Laboratory of Optoelectronic Materials and Technologies and School of Physics, Sun Yat-sen University, Guangzhou 510275, China}
\affiliation{Quantum Engineering Technology Labs, H. H. Wills Physics Laboratory and Department of Electrical \& Electronic Engineering, University of Bristol, BS8 1FD,UK.}
\author{Jianwei Wang}
\affiliation{Quantum Engineering Technology Labs, H. H. Wills Physics Laboratory and Department of Electrical \& Electronic Engineering, University of Bristol, BS8 1FD,UK.}
\affiliation{State Key Laboratory for Mesoscopic Physics and Collaborative Innovation Centre of Quantum Matter, School of Physics, Peking University, Beijing 100871, China}
\author{Callum M. Wilkes}
\affiliation{Quantum Engineering Technology Labs, H. H. Wills Physics Laboratory and Department of Electrical \& Electronic Engineering, University of Bristol, BS8 1FD,UK.}
\author{Thomas Loke}
\affiliation{School of Physics, The University of Western Australia, Crawley WA 6009, Australia}
\author{Sean O'Gara}
\author{Laurent Kling}
\author{Graham D. Marshall}
\author{Raffaele Santagati}
\affiliation{Quantum Engineering Technology Labs, H. H. Wills Physics Laboratory and Department of Electrical \& Electronic Engineering, University of Bristol, BS8 1FD,UK.}
\author{Timothy C. Ralph}
\affiliation{Centre for Quantum Computation and Communication Technology, School of Mathematics and Physics, University of Queensland, Brisbane, Queensland 4072, Australia}
\author{Jingbo B. Wang}
\affiliation{School of Physics, The University of Western Australia, Crawley WA 6009, Australia}
\author{Jeremy L. O'Brien}
\author{Mark G. Thompson}
\author{Jonathan C. F. Matthews}
\email{jonathan.matthews@bristol.ac.uk}
\affiliation{Quantum Engineering Technology Labs, H. H. Wills Physics Laboratory and Department of Electrical \& Electronic Engineering, University of Bristol, BS8 1FD,UK.}
\date{January 30, 2018}

\begin{abstract}

\noindent Integrated optics is an engineering solution proposed for exquisite control of photonic quantum information. Here we use silicon photonics and the linear combination of quantum operators scheme to realise a fully programmable two-qubit quantum processor. The device is fabricated with readily available CMOS based processing and comprises four {nonlinear} photon-sources, four filters, eighty-two beam splitters and fifty-eight individually addressable phase shifters. To demonstrate performance, we programmed the device to implement ninety-eight various two-qubit unitary operations ({with} average quantum process fidelity of 93.2$\pm$4.5\%), a two-qubit quantum approximate optimization algorithm and efficient simulation of Szegedy directed quantum walks. This fosters further use of the linear combination architecture with silicon photonics for future photonic quantum processors.
\end{abstract}

\maketitle

\noindent 
The range and quality of control that a device has over quantum physics determines the extent of quantum information processing (QIP) tasks that it can perform. One device capable of performing any given QIP task is an ultimate goal~\cite{la-nat-464-45} and silicon quantum photonics~\cite{si-jstqe-22-390} has attractive traits to achieve this: photonic qubits are robust to environmental noise~\cite{o2009photonic}, single qubit operations can be performed with high precision~\cite{wilkes2016high}, a high density of reconfigurable components have been used to manipulate coherent light~\cite{sun2013large, harris2017quantum} and established fabrication processes are CMOS compatible. However, quantum control needs to include entangling operations to be relevant to QIP --- this is recognised as one of the most challenging tasks for photonics because 
of the extra resources required for each entangling step~\cite{knill2001scheme,o2009photonic}. Here, we demonstrate a programmable silicon photonics chip that generates two photonic qubits, on which it then performs arbitrary two-qubit untiary operations, including arbitrary entangling operations. This is achieved by using silicon photonics to reach the complexity required to implement an iteration of the linear combination of unitaries architecture~\cite{zhou2011adding,gui2006general} that we have adapted to realise universal two-qubit processing. The device's performance shows that the design and fabrication techniques used in its implementation work well with the linear combination architecture and can be used to realise larger and more powerful photonic quantum processors.

Miniaturisation of \nohl{quantum-}photonic experiments into chip-scale waveguide circuits started~\cite{politi2008silica} from the need to realise many-mode devices with inherent sub-wavelength stability for generalised quantum-interference experiments, such as multi-photon quantum walks~\cite{peruzzo2010quantum} and boson sampling~\cite{spring2013boson,tillmann2013experimental,crespi2013integrated}. Universal six-mode linear optics implemented with a silica waveguide chip (coupled to free-space photon sources and fibre-coupled detectors) demonstrated the principle that single photonic devices can be configured to perform any given linear optics task~\cite{carolan2015universal}. Silicon waveguides promise even greater capability for large-scale photonic processing, because of their third order nonlinearity that enables photon pair generation within integrated structures~\cite{sharping2006generation}, their capacity for integration with single photon detectors~\cite{fa-natcomm-6-5873} and their component density can be more than three orders of magnitude higher than silica~\cite{si-jstqe-22-390}.

\begin{figure*}[htbp!] 
\centering    
\includegraphics[width=0.95\textwidth]{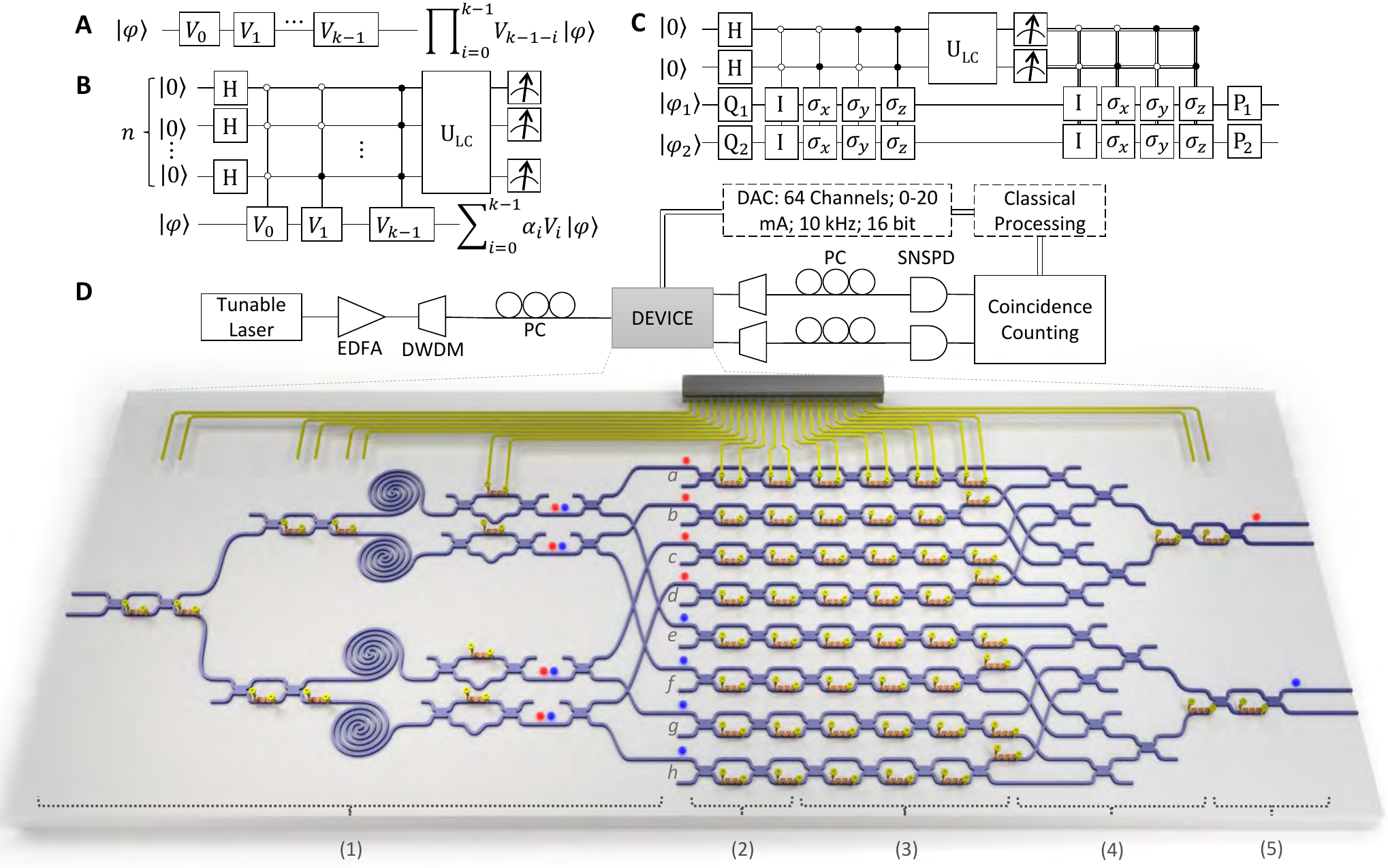}
\caption[Uni_2QP]{\textbf{Quantum information processing circuits and a schematic of the experimental setup.} (A) The conventional quantum circuit model of QIP, that is a multiplication of quantum logic gates in series. 
(B) Probabilistic linear-combination of quantum gates. 
{The operation $\sum_{i=0}^{k-1} {\alpha_i V_i}$ is implemented when all $n$ control qubits are measured to be 0. $U_{\text{LC}}$ is a unitary operation with first row in its matrix representation given as 
$\{ \alpha_0, \alpha_1, \cdots, \alpha_{n-1} \}$ where $\sum_{i=0}^{k-1} |\alpha_i|^2 = 1$, $k=2^n$ and the success probability is $1/k$. 
Other rows are chosen accordingly to make $U_\text{LC}$ unitary. 
} 
{(C) Deterministic linear-combination circuit for universal two-qubit unitary operation. For a $U \in \text{SU}(4)$ being decomposed as Equation (1), $U_{\text{LC}}$ is defined as $\left[\alpha_0, \alpha_1, \alpha_2, \alpha_3; \alpha_1, \alpha_0, -\alpha_3, -\alpha_2; \alpha_2, -\alpha_3, \alpha_0, -\alpha_1; \alpha_3, -\alpha_2, -\alpha_1, \alpha_0 \right]$. The required two auxiliary control qubits can also be replaced by a four-level ququard and then $U_{\text{LC}}$ is a single-ququard operation.} (D) Schematic of our device and external setup. A tunable continuous wave laser is amplified with an optical fibre amplifier (EDFA), spectrally filtered by a dense wavelength-division multiplexing (DWDM) module and launched into the device through a V-groove fibre array (VGA). Photons emerging from the device are collected by the same VGA and two DWDMs are used to separate the signal (red) and idler (blue) photons. Photons are detected by two fibre-coupled superconducting nanowire single-photon detectors (SNSPD). The polarisations of input/output photons are optimised by in-line polarization controllers (PC). Coincidence counting logic records the two-photon coincidence events. Phase shifters on the device are configured through a digital-to-analog converter (DAC), controlled from a computer. The device includes five functional parts: (1) generating ququard-entanglement; (2) preparing initial single-qubit states; (3) implementing single-qubit operations; (4) realizing linear-combination; (5) performing measurement.
}
\label{fig:setup}
\end{figure*}

Programmable quantum processors have been reported with up to five trapped-ion qubits~\cite{de-nat-536-63}, eleven NMR qubits~\cite{vandersypen2001experimental} and tens of superconducting qubits~\cite{so-arxiv:1703.10302}. However, for photons, \nohl{up to} two sequential two-qubit entangling operations implemented with free-space optics~\cite{martin2012experimental, barz2014two} \nohl{and silicon quantum photonics}~\cite{wang2017experimental,2040-8986-19-11-114006} is the state of the art in qubit control. But the degree of control and utility of these photonic demonstrators is limited intrinsically because arbitrary two-qubit processing requires \nohl{the equivalent of} three consecutive entangling gates \nohl{in the circuit model of quantum computing}, as demonstrated experimentally in 2010 with ion-trap quantum processing~\cite{ha-natphys-6-13}. Effective QIP with three sequential entangling operations is beyond the level of complexity that can be practically constructed and maintained with free-space quantum optics or a hybrid of free-space nonlinear optics and integrated linear optics~\cite{carolan2015universal}.

\nohl{Our scheme realizes arbitrary two-qubit unitary operation via a linear combination of four easy-to-implement unitaries --- each being a tensor product of two single-qubit unitaries. The presented chip constructs and exploits high-dimensional entanglement in order to implement the equivalent capability of three sequential entangling gates in the circuit model whilst using only two photons.} It performs universal two-qubit processing with high fidelity whilst all thermal phase shifters in the device are simultaneously active, it is inherently stable and repeatable under continuous operation and it can be reprogrammed at kilohertz rate. We demonstrate the chip's performance by performing process tomography on 98 implemented two-qubit quantum logic gates, by realising the quantum approximate optimization algorithm (QAOA)~\cite{farhi2014quantum,farhi2014quantumapplied} applied to three example constraint satisfaction problems, and by simulating Szegedy quantum walks (SQW)~\cite{szegedy2004spectra,szegedy2004quantum} over an example two-node {weighted} graph. All together, the results presented required 98480 experiment configurations.\\

\noindent \textbf{1. Linear combination of unitaries on a chip for QIP.} 
The conventional quantum circuit model for QIP is a sequence of quantum gates (Fig.~\ref{fig:setup}(A)). The linear combination of unitary operations is an alternative approach (Fig.~\ref{fig:setup}(B)) that is central to various {QIP} tasks~\cite{zhou2011adding,Childs2012Hamiltonian,childs2017quantum,patel2016quantum,wei2016duality,qiang2017quantum}. A universal two-qubit unitary $U \in \text{SU}(4)$ can be implemented by the four-operator linear combination~\cite{SM}
\begin{align}
\textstyle U =\sum_{i=0}^{3}{\alpha_i \left(P_1\sigma_i Q_1\right) \otimes \left(P_2\sigma_i Q_2\right)},
\label{eq:KAKmaintext}
\end{align}
where $P$ and $Q$ are single-qubit gates, $\sigma_i$ are identity and Pauli gates ($I$, $\sigma_x$, $\sigma_y$, $\sigma_z$) and $\alpha_i$ are complex coefficients satisfying $\sum_{i=0}^{3} |\alpha_i|^2 = 1$. 
This linear combination can immediately be implemented through two-qubit version of the $n$-qubit circuit shown in Fig.~\ref{fig:setup}(B), with an intrinsic success probability of 1/4. However, we also note that a deterministic implementation of the linear-combination of $U$ can in principle be achieved with extra classical controlled gates~\cite{SM}, as shown in Fig.~\ref{fig:setup}(C). In the presented chip, \nohl{the linear decomposition of $U$} is implemented probabilistically by expanding the dimension of qubits into qudits and using pre-entanglement between qudit systems that can be generated from parametric photon pair generation~\cite{zhou2011adding,wang2017experimental}. This Hilbert-space-expansion approach \nohl{implements} arbitrary two-qubit unitar\nohl{ies} using resources of only a two-photon entangled-ququard state and mult-mode interferometry, that is inherently stable on our chip~\cite{SM}.

Fig.~\ref{fig:setup}(D) illustrates the schematic of our silicon photonic chip operated with external electrical control, laser input and fibre coupled superconducting detectors. The 7.1 mm $\times$ 1.9 mm chip consists of four spiral-waveguide spontaneous four-wave mixing (SFWM) photon-pair sources~\cite{silverstone2014chip}, four laser pump rejection filters, eighty-two {multi-mode interferometer (MMI) beam splitters} and fifty-eight simultaneously running thermo-optic phase shifters~\cite{silverstone2014chip}. Within the device, the four {SFWM} sources are used to create possible (signal-idler) photon pairs when pumped with a laser that is launched into the chip and split across the four sources according to complex coefficients $\alpha_i$. The spatially bunched photon pairs are coherently generated in either one of the four sources. Post-selecting when signal and idler photons exit at the top two output modes (qubit 1) and the bottom two (qubit 2) respectively, yields a path-entangled ququard state $\ket{\Phi}$ as
\begin{align}
\!\alpha_0\! \ket{1}_a\!\ket{1}_e \!+\!\alpha_1\! \ket{1}_b\ket{1}_f \!+\! \alpha_2\!\ket{1}_c\ket{1}_g\!+\! \alpha_3\!\ket{1}_d\ket{1}_h 
\label{eq:PathEntangledState} 
\end{align} 
at the end of stage (1) marked on the device shown in Fig.~\ref{fig:setup}(D), with intrinsic success probability of {1/4}. $\ket{1}_j$ represents the Fock state in spatial modes labeled by $j=a, b, c, d, e, f, g, h$.

Spatial modes $a$-$h$ are each extended into two modes to form path-encoded qubits $\ket{\varphi_1}$ or $\ket{\varphi_2}$ with arbitrary amplitude and phase controlled with Mach Zehnder Interferometer\nohl{s} (MZI) and an extra phase shifter. Single-qubit operations $A_i$ (=$P_1\sigma_i Q_1$) and $B_i$ (=$P_2\sigma_i Q_2$) are applied using MZIs and phase shifters to $\ket{\varphi_1}$ and $\ket{\varphi_2}$ respectively, evolving $\ket{\Phi}$ into
\begin{align}
\textstyle \sum_{i=0}^{3}\alpha_i A_i\ket{\varphi_1}_{u^i}B_i\ket{\varphi_2}_{v^i}, 
\end{align}
where $u^i \in \{a,b,c,d\}$ and $v^i \in\{e,f,g,h\}$. Next, the qubits $a,b,c,d$ are combined into one final-stage qubit, and the qubits $e,f,g,h$ into the remaining final-stage qubit, as shown in stage (4) of Fig.~\ref{fig:setup}(D) with intrinsic success probability {$1/16$}. This removes path information of the signal (idler) photon and thus we obtain the final evolved two-photon state as 
\begin{align}
\textstyle \left(\sum_{i=0}^{3}{\alpha_i A_{i}\otimes B_{i}} \right) \ket{\varphi_1}\ket{\varphi_2}.
\end{align}
Once photons are generated, the overall intrinsic success probability of our design is $1/64$, which is higher than the two main schemes considered for universal linear optical quantum computation~\cite{SM}: the Knill-Laflamme-Milburn (KLM) scheme~\cite{knill2001scheme} and linear optical measurement-based quantum computation (MBQC)~\cite{raussendorf2001one}. The success probability of this optical implementation could be further increased to $1/4$ if we were to separate signal and idler photons with certainty and use also an advanced linear-combination circuit that utilizes the unused optical ports in our current chip design~\cite{SM}.\\

\noindent\textbf{2. Realising individual quantum gates.} The linear-combination scheme can simplify implementation of families of two-qubit gates. For example, an arbitrary two-qubit controlled-unitary gate CU can be implemented as the linear combination of two terms:
\begin{align}
\textstyle \text{CU} = \frac{1}{\sqrt{2}}\begin{pmatrix} 1 & 0\\ 0 & i \end{pmatrix} \otimes \frac{I-iU}{\sqrt{2}} + \frac{1}{\sqrt{2}}\begin{pmatrix} 1 & 0\\ 0 & -i \end{pmatrix} \otimes \frac{I+iU}{\sqrt{2}},
\end{align}
and SWAP gate can be implemented by a linear combination of only identity and Pauli gates:
\begin{align}
&\text{SWAP} = \frac{1}{2}\left( I\otimes I + \sigma_x\otimes \sigma_x + \sigma_y\otimes \sigma_y + \sigma_z\otimes \sigma_z  \right).
\end{align} 

\begin{figure}[htbp!] 
\centering    
\includegraphics[width=0.48\textwidth]{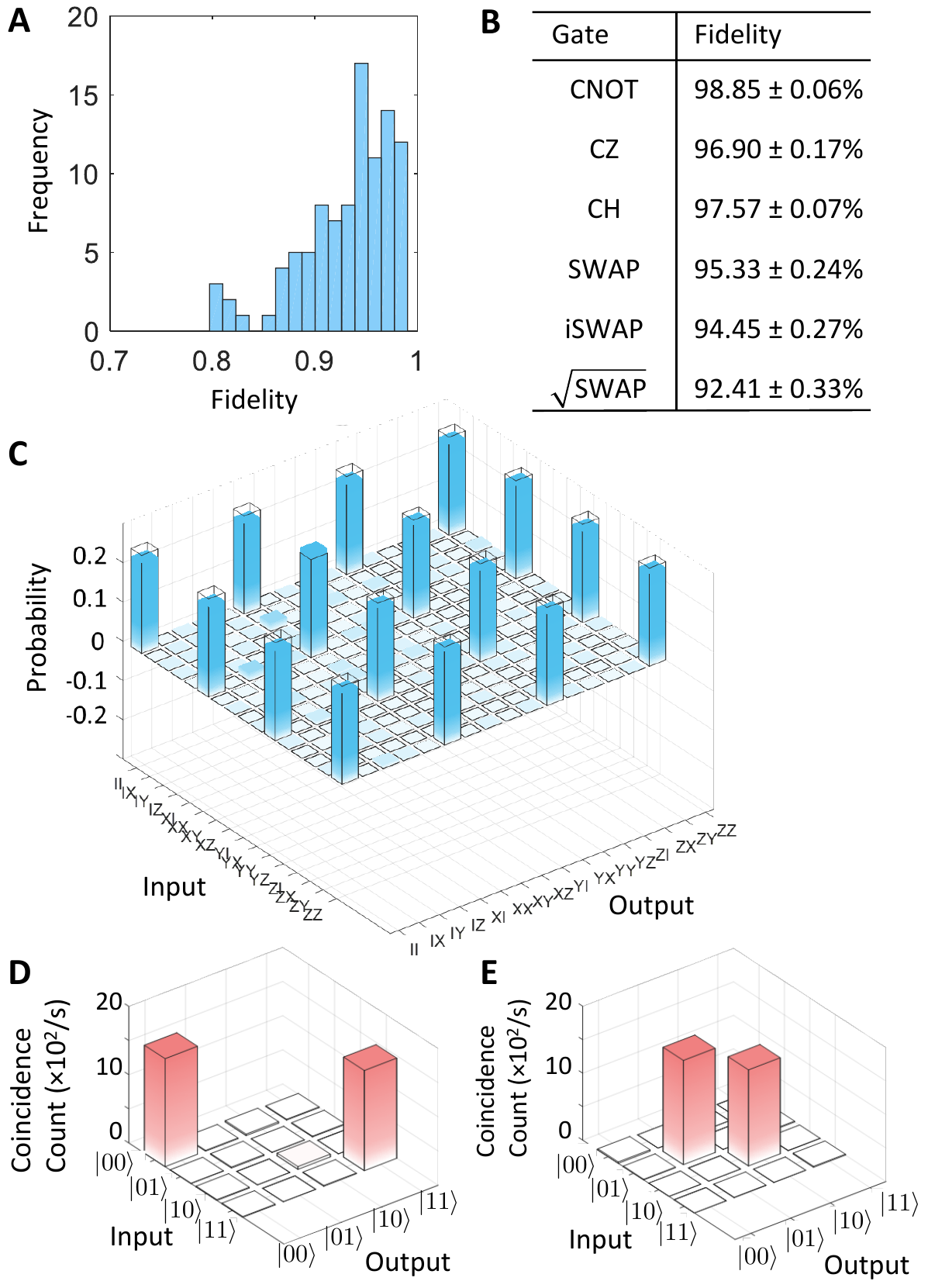}
\caption[DataAll]{\textbf{Experimental realisation of arbitrary 2-qubit gates.} (A) A histogram of measured process fidelities for 98 two-qubit quantum gates ($\bar{F}=93.15\pm 4.53\%$). (B) Measured process fidelities for example two-qubit gates: C-NOT, C-Z, C-H, SWAP, iSWAP, $\sqrt{\text{SWAP}}$. (C) The real part of experimentally determined process matrices of SWAP, with ideal theoretical values overlaid. (D, E) Logical basis truth tables for entanglement filter (D) and entanglement splitter (E).  
}
\label{fig:ExpData}
\end{figure}

To show the reconfigurability and performance of our chip, we implemented 98 different two-qubit quantum logic gates, for which we performed on-chip full quantum process tomography and reconstructed the process matrix using the maximum likelihood estimation technique for each gate~\cite{SM}. A histogram of measured process fidelities for these 98 gates is shown in Fig.~\ref{fig:ExpData}(A), with a mean statistical fidelity of 93.15$\pm$4.53\%. The implemented gates include many common instances---as shown in Fig.~\ref{fig:ExpData}(B, C)---achieving high fidelities, such as CNOT with 98.85$\pm$0.06\% and SWAP with 95.33$\pm$0.24\%. Our device also allows implementation of non-unitary quantum operations. The entanglement filter (EF)~\cite{okamoto2009entanglement,zhou2011adding} and the entanglement splitter (ES)~\cite{zhou2011adding} can be implemented by
\begin{align}
\text{EF} = (I\otimes I + \sigma_z\otimes \sigma_z)/\sqrt{2} \\
\text{ES} = (I\otimes I - \sigma_z\otimes \sigma_z)/\sqrt{2}
\end{align} 
The results are shown in Fig.~\ref{fig:ExpData}(D) and (E) in the form of logical basis truth tables, with the classical fidelities of 95.31$\pm$0.45$\%$ and 97.69$\pm$0.31$\%$ respectively.\\

\begin{figure}[htbp!] 
\centering    
\includegraphics[width=0.5\textwidth]{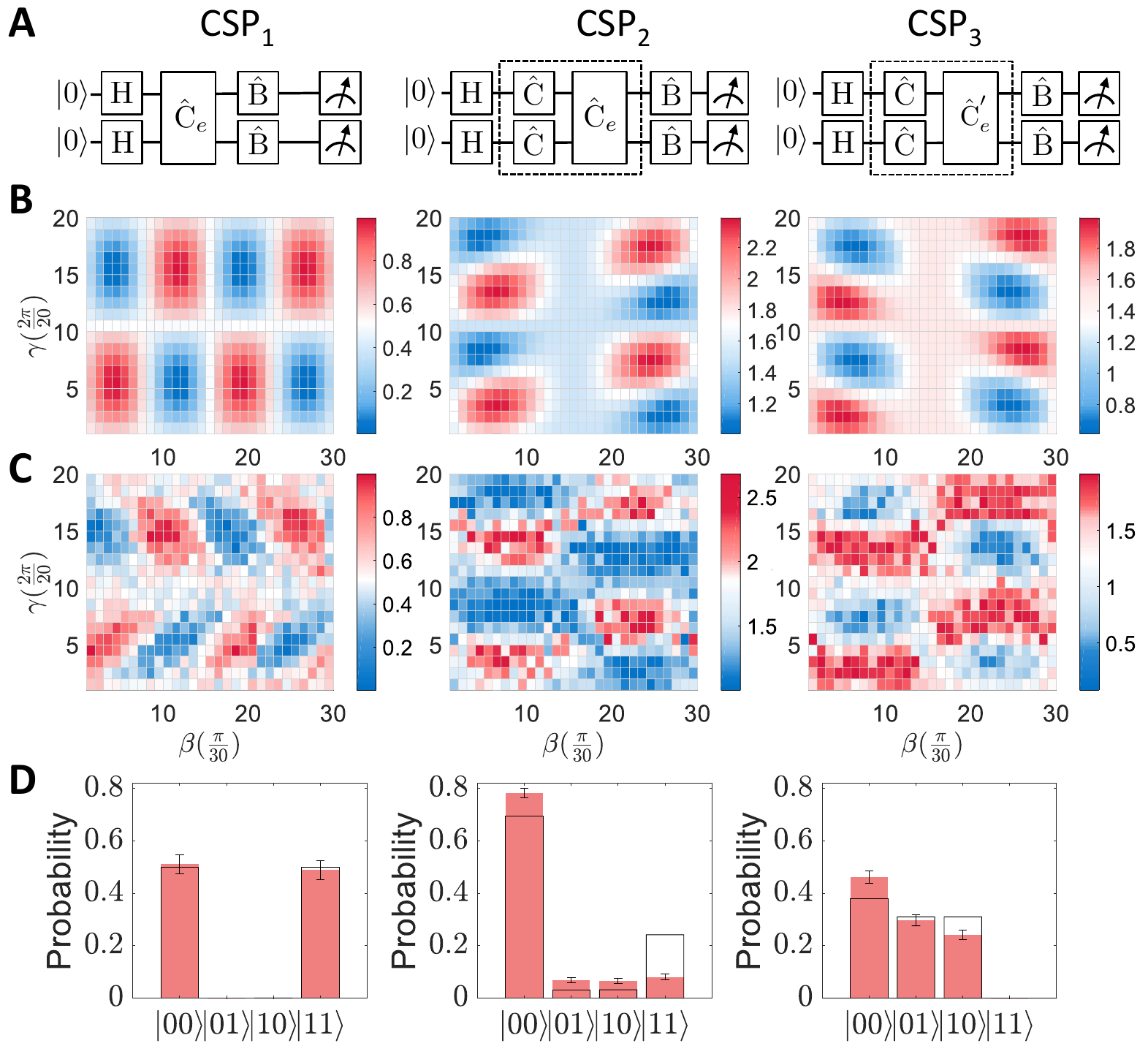}
\caption[DataAll]{
\textbf{Experimental realisation of a two-qubit quantum approximate optimisation algorithm.} Panels arranged into three columns, corresponding to three example CSPs, labeled 1-3.
(A) Quantum circuits of QAOA for each CSP. (B) Theoretical and (C) experimentally determined values of $\braket{{\gamma},{\beta}|C|{\gamma},{\beta}}$ over the grid of $[\gamma,\beta]\in[0, 2\pi] \times [0, \pi]$ for $\text{CSP}_1$, $\text{CSP}_2$ and $\text{CSP}_3$, with step size $\delta_{\gamma}=\frac{2\pi}{20},\delta_{\beta}=\frac{\pi}{30}$, for finding the optimized $\ket{\gamma,\beta}$ states. (D) Experimental measurement results of the optimized $\ket{\gamma,\beta}$ states, outputting the searched target string $z$ for each CSP 1-3. }
\label{fig:ExpData3}
\end{figure}

\begin{figure*}[htbp!] 
\centering    
\includegraphics[width=1.0\textwidth]{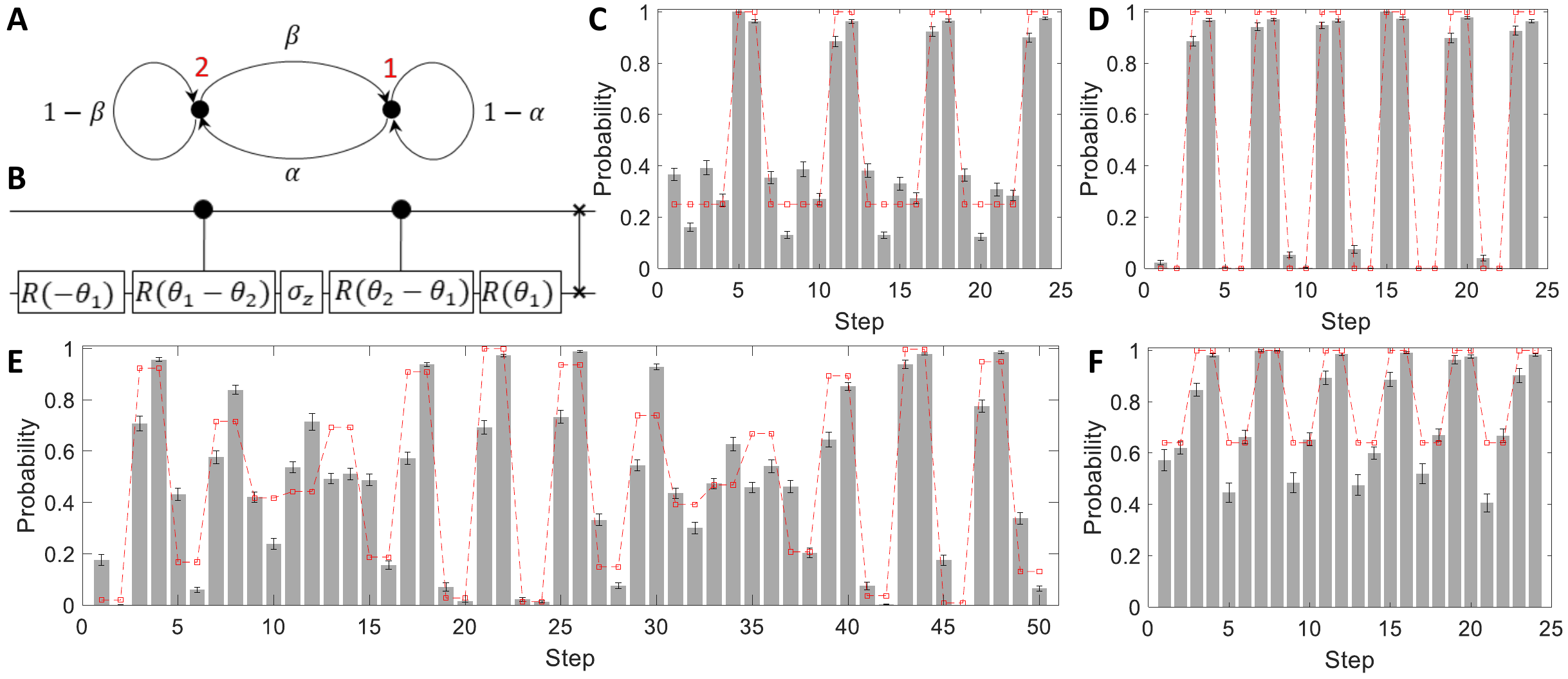}
\caption[DataAll]{
\textbf{Experimental quantum simulation of Szegedy directed quantum walks.}
(A) A weighted two-node graph. Edge weights are decided by $\alpha, \beta$ $\in [0,1]$. (B) Quantum circuit for a single-step SQW on the two-node graph. $R(\theta)$ is defined as $R(\theta) = \left[\cos{\theta}, -\sin{\theta};\sin{\theta},\cos{\theta}\right]$ with $\theta \in \{\theta_1, -\theta_1, (\theta_1-\theta_2), (\theta_2 - \theta_1)\}$ where $\theta_1 =\text{arccos}(\sqrt{1-\alpha})$ and $\theta_2=\text{arccos}(\sqrt{\beta})$. 
(C-F) Theoretical (red points) and experimental (gray bars) probability distributions (of the walker being at node 1) of SQWs with the initial state $\ket{00}$: (C) $\alpha=\beta=0.25$, $F_\text{avg}=98.46\pm0.04\%$; (D) $\alpha=\beta=0.5$, $F_\text{avg}=98.48\pm0.04\%$; (E) $\alpha=\beta=0.43$, $F_\text{avg}=98.02\pm 0.04\%$; (F) $\alpha = 0.1, \beta=0.9$, $F_\text{avg}=98.35\pm0.15\%$. }
\label{fig:ExpData4}
\end{figure*}

\noindent\textbf{3. Implementing a two-qubit Quantum Approximate Optimization Algorithm for Constraint Satisfaction Problems.} 
The quantum approximate optimization algorithm (QAOA) was proposed for finding approximate solutions to combinatorial search problems such as constraint satisfaction problems (CSPs)~\cite{farhi2014quantum,farhi2014quantumapplied}. It is a promising candidate to run on primitive quantum computers because of its possible use for optimization and its conjectured potential as a route to establishing quantum supremacy~\cite{farhi2016quantum}. \nohl{A general CSP is specified by $n$ bits and a collection of $m$ constraints---each of which involves a small subset of the bits. For a CSP, QAOA outputs a binary string $z$ which (approximately) maximizes the number of satisfied constraints, i.e., $C(z)=\sum_{l=1}^{m}C_l(z)$ where $C_{l}(z)=1$ if $z$ satisfies the $l$-th constraint, otherwise 0 --- this is the goal of CSP.}

The QAOA process can be summarised as follows. Suppose two operators $C$ and $B$ are defined as 
\begin{align}
C\ket{z}:=C(z)\ket{z},~B:= \textstyle \sum_{i=1}^n{\sigma_x^{(i)}}
\end{align}
where $\sigma_x^{(i)}$ represents $\sigma_x$ acting on the $i$-th qubit, and a quantum state $\ket{\vec{\gamma},\vec{\beta}}$ is defined as 
\begin{align}
\ket{\vec{\gamma},\vec{\beta}}=e^{-i\beta_pB}e^{-i\gamma_pC}\cdots e^{-i\beta_1B}e^{-i\gamma_1C}H^{\otimes n}\ket{0}^{\otimes n}
\end{align}
where $\vec{\gamma}:=(\gamma_1,\cdots, \gamma_p) \in [0, 2\pi]^p$ and $\vec{\beta}:=(\beta_1,\cdots, \beta_p) \in [0, \pi]^p$. QAOA seeks the target string $z$ by searching the $\vec{\gamma}$ and $\vec{\beta}$ that maximize $\braket{\vec{\gamma},\vec{\beta}|C|\vec{\gamma},\vec{\beta}}$ and then the corresponding state $\ket{\vec{\gamma},\vec{\beta}}$ in the computational basis \nohl{encodes the solution}. For a given $\vec{\gamma}$ and $\vec{\beta}$, $\braket{\vec{\gamma},\vec{\beta}|C|\vec{\gamma},\vec{\beta}}$ can be evaluated through a quantum computer, which can further be used as a subroutine in an enveloping classical algorithm---for example, run the quantum computer with angles ($\vec{\gamma}$, $\vec{\beta}$) from a {fine} grid on the set $[0,2\pi]^p\times[0,\pi]^p$---to find the best $\vec{\gamma}$ and $\vec{\beta}$ for maximizing $\braket{\vec{\gamma},\vec{\beta}|C|\vec{\gamma},\vec{\beta}}$~\cite{farhi2016quantum}. With $p$ getting increased, the quality of the approximation of QAOA improves~\cite{farhi2014quantum}.

In our experiments, we restricted to the $p=1$ case of QAOA, and applied QAOA to three 2-bit CSPs. The corresponding quantum circuits are shown in Fig.~\ref{fig:ExpData3}(A). The first CSP {(denoted as ${\text{CSP}}_1$) is the 2-bit Max2Xor problem which has just one constraint as $C(z)=\frac{1}{2}+\frac{1}{2}z_1 z_2$ where $z_1, z_2 \in \{\pm1\}$. The other two CSPs have three constraints:
\begin{align}
&\text{CSP}_2:  \\ 
&\textstyle  C_1(z) = \frac{1}{2}+\frac{1}{2}z_1;  C_2(z) = \frac{1}{2}+\frac{1}{2}z_2;  C_3(z) = \frac{1}{2} + \frac{1}{2}z_1z_2 \nonumber \\
&\text{CSP}_3: \\
&\textstyle  C_1(z) = \frac{1}{2}+\frac{1}{2}z_1;  C_2(z) = \frac{1}{2}+\frac{1}{2}z_2;  C_3(z) = \frac{1}{2} - \frac{1}{2}z_1z_2 \nonumber
\end{align}
For the $p=1$ QAOA, there are only two angles, $\gamma$ and $\beta$, to be found for optimizing $\braket{\gamma, \beta|C|\gamma,\beta}$. We search $\gamma$ and $\beta$ along a fine grid on the compact set $[0, 2\pi] \times [0, \pi]$ and show each obtained value of $\braket{\gamma, \beta|C|\gamma,\beta}$ as in Fig.~\ref{fig:ExpData3}(B,C) where the target angles are marked as the reddest block. By measuring the corresponding $\ket{\gamma,\beta}$ state in the computational basis for $\text{CSP}_1$, we obtain ``00'' or ``11'' with highest probability, corresponding to the target string of $\text{CSP}_1$: $\{z_1,z_2\}=\{1,1\}$ or $\{-1,-1\}$. Similarly, the obtained results for $\text{CSP}_2$ is $\{z_1,z_2\}=\{1,1\}$ and for $\text{CSP}_3$ are $\{z_1,z_2\}=\{1,1\}$, $\{1,-1\}$ or $\{-1,1\}$---either of which is a solution of $\text{CSP}_3$. The experimental results are shown in Fig.~\ref{fig:ExpData3}(D), with the classical fidelities between experiment and theory of 99.88$\pm$0.10\%, 96.98$\pm$0.56\% and 99.48$\pm$0.27\% respectively.

\noindent\textbf{4. Simulating Szegedy Quantum Walks.} Quantum walks model a quantum particle's random movement in a discretized space according to a given set of rules known as a graph. They are of interest for developing quantum computing (e.g. Ref.~\onlinecite{childs2013universal}) and quantum algorithms (e.g. Ref.~\onlinecite{childs2004spatial}) and as an observable quantum phenomena~\cite{peruzzo2010quantum}. The Szegedy quantum walk (SQW)~\cite{szegedy2004spectra,szegedy2004quantum} is a \nohl{particular} class that allows unitary evolution on directed and weighted graphs---which the standard discrete-time and continuous-time quantum walk formalisms do not permit---and has been proposed for application to quantum speedup for ranking the relative importance of nodes in connected database~\cite{paparo2011google,paparo2013quantum,loke2017comparing}. The realization of SQW-based algorithms on a quantum computer requires an efficient quantum circuit implementation for the walk itself~\cite{chiang2009efficient,loke2017efficient}. Here we have implemented SQW experimentally on an example two-node graph.

A general weighted graph $G$ {with $N$ nodes} can be described by its transition matrix $P$ where an element $P_{i,j}$ is given by the weight of a directed edge from node $i$ to $j$, satisfying $\sum_{i=0}^{N-1}{P_{i,j}}=1$. A SQW on $G$ is defined as a discrete-time unitary time evolution on a Hilbert space $H=H_1\otimes H_2$ where $H_1$ and $H_2$ are both $N$-dimensional Hilbert spaces. The single-step operator of an SQW is given by $U_{\text{sz}} = S(2\Pi - I)$. Here $S$ is a SWAP operator defined as $S=\sum_{i=0}^{N-1}\sum_{j=0}^{N-1} {\ket{i,j}\bra{j,i}}$, and $\Pi$ is a projection operator as $\Pi = \sum_{i=0}^{N-1}{\ket{\phi_i}\bra{\phi_i}}$ with $\ket{\phi_i}=\ket{i}\otimes \sum_{j=0}^{N-1}{\sqrt{P_{j+1,i+1}}\ket{j}}$ for $i \in \{0, \cdots, N-1\}$. For the example two-node graph that we label $\mathcal{E}$ and sketched in Fig.~\ref{fig:ExpData4}(A), a quantum circuit implementation for single-step SQW operator can be constructed by using the scheme proposed in ref~\cite{loke2017efficient}, as shown in Fig.~\ref{fig:ExpData4}(B). Repeating this circuit generates an efficient quantum circuit implementation of multiple-step SQWs, which holds exponential speedup over classical simulations~\cite{loke2017efficient}.

The periodicity of SQWs is determined by the eigenvalues of the single-step operator $U_{\text{sz}}$, and it has been studied on several families of finite graphs~\cite{higuchi2017periodicity}. $U_{\text{sz}}$ of the graph $\mathcal{E}$ has four eigenvalues: $\{-1, 1, 1-s-\sqrt{s^2 -2s}, 1-s+\sqrt{s^2-2s} \}$ where $s=\alpha + \beta$ and $\alpha,\beta \in\left[0,1\right]$. $U_{\text{sz}}$ is periodic if and only if there exists an integer $n$ such that $\lambda_i^n=1$ for all four eigenvalues $\lambda_i$ of $U_{\text{sz}}$. The period is then given by the lowest common multiple of the periods of the eigenvalues. $\mathcal{E}$ has a symmetric transition matrix when $\alpha = \beta$. For SQWs on a symmetric graph $\mathcal{E}$, periodic walks exist in the cases $\alpha=\beta=\frac{1}{4}, \frac{1}{2}, \frac{3}{4}, 1$---with periods of 6, 4, 6 and 2 steps respectively---of which the first two are experimentally verified as shown in Fig.~\ref{fig:ExpData4}(C) and (D). SQWs on a general instance of $\mathcal{E}$ do not exhibit perfect periodicity, as shown by Fig.~\ref{fig:ExpData4}(E) that shows the behaviour of SQWs on $\mathcal{E}$ with $\alpha=\beta=0.43$. $\mathcal{E}$ has an asymmetric transition matrix when $\alpha \ne \beta$, and perfect periodicities of SQWs can exist in particular cases, such as $\alpha+\beta =1$, which has a period of 4 steps. An example of this kind, $\alpha=0.1, \beta=0.9$, is shown in Fig.~\ref{fig:ExpData4}(F). In our device, we can also perform state tomography on a given time-evolved state of SQWs --- we have performed quantum state tomography for more than 500 time-evolved states, observing an average state fidelity of 93.95$\pm$2.52\% with theoretical prediction. 
\\

\noindent\textbf{5. Discussion.} The underlying optical linear-combination protocol used here requires the number of beamsplitters and phase shifters in the optical network to increase exponentially with the number of qubits. Although this scheme is ultimately unscalable, it is suitable in the near- and mid-term for situations where the photonic components are easier to create than the qubits themselves and it is no less demanding of individual component performance than other linear optics approaches to QIP. Our range of demonstrations with a single device has shown that the linear-combination scheme is valuable in permitting QIP demonstrations with the current state of the art in photonics and that silicon photonics is capable of fulfilling its requirements. The device reported comprises nonlinear photon sources, optical filtering and reconfigurable linear optics and it was fabricated with a standard CMOS based silicon photonics processes onto a single photonic chip. It generates photons, encodes quantum information on them, manipulates the\nohl{m} and performs tomographic measurement, all with high fidelity quantum control for thousands of configurations. From our experience, our demonstrations of the QAOA and SQWs are beyond the practicality and performance achievable with free-space bulk optical experiments and glass-based integrated photonics. Together with developing multi-photon sources~\cite{collins2013integrated} and integration with on chip detection~\cite{kh-natpho-10-727}, future \nohl{iterations of silicon photonics} opens the way to more sophisticated photonic experiments that are impossible to achieve otherwise, including \nohl{the eventual} full-scale universal quantum technologies using light~\cite{gi-prl-115-020502}.

\noindent\textbf{Data access statement:} The data that support the findings of this study are available from the corresponding author upon reasonable request.

\clearpage


\pagebreak
\newpage
\onecolumngrid
\appendix
\begin{center}
\section*{Supplementary Information}
\end{center}
\setcounter{equation}{0}
\setcounter{figure}{0}
\setcounter{table}{0}
\setcounter{page}{1}
\makeatletter
\renewcommand{\theequation}{S\arabic{equation}}
\renewcommand{\thefigure}{S\arabic{figure}}
\renewcommand{\bibnumfmt}[1]{[#1]}
\renewcommand{\citenumfont}[1]{#1}
\normalsize

\section{Linear-combination scheme for universal two-qubit unitary operation}

\subsection{Linear-combination decomposition of SU(4)}
By using the Cartan's KAK decomposition~\cite{khaneja2001cartan}, an arbitrary two-qubit unitary operation $U \in \text{SU(4)}$ can be decomposed as:
\begin{align}
U = (P_1 \otimes P_2)U_\text{D}(Q_1 \otimes Q_2), \label{eq:ch6KAK}
\end{align}
where $P_1, P_2, Q_1$ and $Q_2$ are single-qubit gates, and $U_\text{D}$ represents the gate below:
\begin{align}
U_\text{D} = \exp (-i(k_1 \sigma_1 \otimes \sigma_1 + k_2 \sigma_2\otimes \sigma_2 +k_3 \sigma_3 \otimes \sigma_3)).
\end{align}
Here $k_i$ ($i = 1, 2, 3$) are real numbers, and $\sigma_i$ are Pauli matrices $\sigma_x$, $\sigma_y$ and $\sigma_z$. A step-by-step procedure of applying Cartan's KAK decomposition onto $\text{SU(4)}$ can be found in ref~\onlinecite{tucci2005introduction}.

Considering the fact that
\begin{align}
\exp(iAx) = \cos (x)I + i\sin(x)A
\end{align}
for an arbitrary real number $x$ and a matrix $A$ satisfying $A^2 = I$~(ref.~\onlinecite{nielsen2010quantum}), $U_\text{D}$ can be reformed as:
\begin{align}
U_\text{D} & = (\cos(k_1)I\otimes I-i\sin(k_1) \sigma_x \otimes \sigma_x) \cdot (\cos(k_2)I\otimes I-i\sin(k_2)\sigma_y \otimes \sigma_y) \cdot (\cos(k_3)I\otimes I-i\sin(k_3)\sigma_z \otimes \sigma_z)   \label{eq:linearCoefficients}
\end{align}
We also have the results that 
\begin{align}
\sigma_x \sigma_y & = - \sigma_y \sigma_x = i \sigma_z,  \label{eq:pauliOpRelation1}\\
\sigma_y \sigma_z &= - \sigma_z \sigma_y = i \sigma_x, \label{eq:pauliOpRelation2}\\
\sigma_z \sigma_x &= - \sigma_x \sigma_z = i \sigma_y.
\label{eq:pauliOpRelation3}
\end{align}
From Equations~\eqref{eq:ch6KAK}, \eqref{eq:linearCoefficients} and \eqref{eq:pauliOpRelation1}-\eqref{eq:pauliOpRelation3}, we can rewrite $U$ into a linear-combination form:
\begin{align}
U =\alpha_0 A_0 \otimes B_0 + \alpha_1 A_1 \otimes B_1 +\alpha_2 A_2\otimes B_2 +\alpha_3 A_3 \otimes B_3
\label{eq:KAK}
\end{align}
where $\alpha_0$, $\alpha_1$, $\alpha_2$ and $\alpha_3$ are four complex coefficients obtained from $k_i$:
\begin{align}
\alpha_0 & = (\cos(k_1)\cos(k_2)\cos(k_3) - i\sin(k_1)\sin(k_2)\sin(k_3)) ,  \nonumber \\
\alpha_1 & =( \cos(k_1)\sin(k_2)\sin(k_3) - i \sin(k_1)\cos(k_2)\cos(k_3)),\nonumber \\
\alpha_2 & =(\sin(k_1)\cos(k_2)\sin(k_3) - i \cos(k_1)\sin(k_2)\cos(k_3) ), \nonumber \\
\alpha_3 & =(\sin(k_1)\sin(k_2)\cos(k_3) - i\cos(k_1)\cos(k_2)\sin(k_3) ),  \label{linearCoefficients}
\end{align}
and they satisfy $\sum_{i=0}^{3} |\alpha_i|^2 = 1$. It is easy to verify that $A_i$, $B_i$ ($i=0,\cdots, 3$) are single-qubit operations as below:
\begin{align}
A_0 & = P_1 I Q_1,~~~~B_0 = P_2 I Q_2, \nonumber \\
A_1 & = P_1 \sigma_x Q_1,~~~B_1 = P_2 \sigma_x Q_2, \nonumber \\
A_2 & = P_1 \sigma_y Q_1,~~~B_2 = P_2 \sigma_y Q_2, \nonumber \\
A_3 & = P_1 \sigma_z Q_1,~~~B_3 = P_2 \sigma_z Q_2. \label{eq:8Ops}
\end{align}

\subsection{Probabilistic circuit for linear-combination of quantum gates}
Here we present more details of the circuit shown in Fig. 1(B) in the main text that is the design to implement the general linear combination of quantum gates probabilistically. $U_{\text{LC}}$ in the circuit is a unitary operation defined as
\begin{align}
U_{\text{LC}}=\begin{pmatrix} \alpha_0& \alpha_1& \cdots& \alpha_{n-1}\\ u_{2,1}& u_{2,2}& \cdots& u_{2,n-1}\\ \vdots& \vdots& \ddots& \vdots \\ u_{n-1,1}& u_{n-1,2}&\cdots& u_{n-1,n-1} \end{pmatrix},
\end{align}
whose first row decides the linear coefficients $\alpha_i$.
The circuit evolves as below:
\begin{align}
&\overbrace {\ket{00 \cdots 0}}^k\ket{\varphi} \\
\to &\frac{1}{2^{k/2}} \left( \overbrace {\ket{00 \cdots 0}}^k\ket{\varphi} + \overbrace {\ket{00 \cdots 1}}^k\ket{\varphi} + \cdots + \overbrace {\ket{11 \cdots 1}}^k\ket{\varphi}  \right) \\
\to &\frac{1}{2^{k/2}} \left( \overbrace {\ket{00 \cdots 0}}^k V_0\ket{\varphi} + \overbrace {\ket{00 \cdots 1}}^k V_1\ket{\varphi} + \cdots + \overbrace {\ket{11 \cdots 1}}^k V_{n-1}\ket{\varphi}  \right) \\
\to &\frac{1}{2^{k/2}} \left[ \left(\alpha_0 \overbrace {\ket{00 \cdots 0}}^k + u_{2,1}\overbrace {\ket{00 \cdots 1}}^k + \cdots + u_{n-1,1}\overbrace {\ket{11 \cdots 1}}^k \right) V_0 \ket{\varphi} \right. \\ 
&~~~~~\left. + \left(\alpha_1 \overbrace {\ket{00 \cdots 0}}^k  + u_{2,2}\overbrace {\ket{00 \cdots 1}}^k  + \cdots + u_{n-1,2}\overbrace {\ket{11 \cdots 1}}^k\right) V_1\ket{\varphi} \right. \\
&~~~~~\left. \cdots \right. \\
&~~~~~\left. +\left(\alpha_{n-1} \overbrace {\ket{00 \cdots 0}}^k + u_{2,n-1}\overbrace {\ket{00 \cdots 1}}^k + \cdots + u_{n-1,n-1}\overbrace {\ket{11 \cdots 1}}^k \right) V_{n-1}\ket{\varphi} \right] \\
\to & \frac{1}{2^{k/2}} \left[ \overbrace {\ket{00 \cdots 0}}^k \left(\sum_{i=0}^{n-1} {\alpha_i V_i}\right) \ket{\varphi} + \overbrace {\ket{00 \cdots 1}}^k \left(\sum_{i=0}^{n-1} {u_{2,i} V_i}\right) \ket{\varphi} + \cdots + \overbrace {\ket{11 \cdots 1}}^k \left(\sum_{i=0}^{n-1} {u_{n-1,i} V_i}\right) \ket{\varphi} \right]
\end{align}
When all auxiliary qubits were measured to be ``0'' in the computational basis, the linear-combination operation $\sum_{i=0}^{n-1}{\alpha_i V_i}$ would be implemented. The success probability is $1/k$, where $k=2^n$.

\subsection{Deterministic linear-combination circuit for universal two-qubit unitary gates}
The proposed circuit shown in Fig.~1(C) in the main text can implement the linear combination of four tensor products of two single-qubit gates, i.e., the universal two-qubit unitary gate according to Equation \eqref{eq:KAK} deterministically. Suppose the four linear coefficients are $\alpha_0$, $\alpha_1$, $\alpha_2$ and $\alpha_3$, the required $U_\text{LC}$ in the circuit is given by: 
\begin{align}
U_{\text{LC}} = \begin{pmatrix}
\alpha_0 & \alpha_1 & \alpha_2 & \alpha_3 \\
\alpha_1 & \alpha_0 & -\alpha_3 & -\alpha_2\\
\alpha_2 & -\alpha_3 & \alpha_0 & -\alpha_1 \\
\alpha_3 & -\alpha_2 & -\alpha_1 & \alpha_0
\end{pmatrix}
\label{eq:ULC_2qubit}
\end{align} 

We first show that such a unitary $U_\text{LC}$ always exists for an arbitrary $U \in \text{SU}(4)$. 
\begin{proof}
An arbitrary $U \in \text{SU}(4)$ can be decomposed into the form
\begin{align}
\textstyle U &=\sum_{i=0}^{3}{\alpha_i \left(P_1\sigma_i Q_1\right) \otimes \left(P_2\sigma_i Q_2\right)} = \left(P_1\otimes P_2\right) \left( \sum_{i=0}^{3}{\alpha_i \sigma_i \otimes\sigma_i }\right)\left(Q_1\otimes Q_2\right)
\end{align}
(see Section I.A). $U$ being unitary implies that its non-local part: $U_\text{D} = \sum_{i=0}^{3}{\alpha_i \sigma_i \otimes\sigma_i}$ is unitary, i.e., $U_\text{D}{U_\text{D}}^{\dag}=I\otimes I$. On the other hand, 
\begin{align}
UU^\dag &=\left( \alpha_0 I \otimes I + \alpha_1 \sigma_x \otimes \sigma_x + \alpha_2 \sigma_y \otimes \sigma_y + \alpha_3 \sigma_z \otimes \sigma_z \right) \left(  {\alpha_0}^\dag I \otimes I + {\alpha_1}^\dag \sigma_x \otimes \sigma_x + {\alpha_2}^\dag \sigma_y \otimes \sigma_y + {\alpha_3}^\dag \sigma_z \otimes \sigma_z \right) \\
&=\left( |\alpha_0|^2 +|\alpha_1|^2  +|\alpha_2|^2  +|\alpha_3|^2  \right)I \otimes I +\left(\alpha_0 {\alpha_1}^\dag + {\alpha_0}^\dag \alpha_1 - \alpha_2 {\alpha_3}^\dag -{\alpha_2}^\dag \alpha_3 \right) \sigma_x \otimes \sigma_x \nonumber \\
&~~~+\left( \alpha_0 {\alpha_2}^\dag + {\alpha_0}^\dag \alpha_2 - \alpha_1 {\alpha_3}^\dag -{\alpha_1}^\dag \alpha_3 \right) \sigma_y \otimes \sigma_y + \left( \alpha_0 {\alpha_3}^\dag + {\alpha_0}^\dag \alpha_3 - \alpha_1 {\alpha_2}^\dag - {\alpha_1}^\dag \alpha_2 \right) \sigma_z \otimes \sigma_z
\end{align}
This gives that 
\begin{align}
|\alpha_0|^2 +|\alpha_1|^2  +|\alpha_2|^2  +|\alpha_3|^2 &= 1 \\
\alpha_0 {\alpha_1}^\dag + {\alpha_0}^\dag \alpha_1 - \alpha_2 {\alpha_3}^\dag -{\alpha_2}^\dag \alpha_3 &=0 \\
\alpha_0 {\alpha_2}^\dag + {\alpha_0}^\dag \alpha_2 - \alpha_1 {\alpha_3}^\dag -{\alpha_1}^\dag \alpha_3 & = 0 \\ 
\alpha_0 {\alpha_3}^\dag + {\alpha_0}^\dag \alpha_3 - \alpha_1 {\alpha_2}^\dag - {\alpha_1}^\dag \alpha_2 &=0
\end{align}
which implies that the operation $U_{\text{LC}}$ as defined in Equation~\eqref{eq:ULC_2qubit} is unitary. 
\end{proof}

The circuit evolves as below (ignore $P_1$, $P_2$, $Q_1$ and $Q_2$ first):
\begin{align}
&\ket{00}\ket{\varphi_1}\ket{\varphi_2} \\
\to&\frac{1}{2} \left( \ket{00} + \ket{01} + \ket{10} + \ket{11}  \right)\ket{\varphi_1}\ket{\varphi_2} \\
\to & \frac{1}{2} \left( \ket{00} I \ket{\varphi_1} I\ket{\varphi_2}+ \ket{01}\sigma_x\ket{\varphi_1}\sigma_x\ket{\varphi_2} + \ket{10}\sigma_y\ket{\varphi_1}\sigma_y\ket{\varphi_2} + \ket{11}\sigma_z\ket{\varphi_1}\sigma_z\ket{\varphi_2}  \right) \\
\to & \frac{1}{2} \left[ \left( \alpha_0 \ket{00} + \alpha_1 \ket{01} + \alpha_2 \ket{10} + \alpha_3 \ket{11} \right) I \ket{\varphi_1}  I\ket{\varphi_2} + \left( \alpha_1 \ket{00} + \alpha_0 \ket{01} - \alpha_3 \ket{10} - \alpha_2 \ket{11} \right) \sigma_x \ket{\varphi_1} \sigma_x \ket{\varphi_2} \right. \nonumber \\
&~~\left. + \left( \alpha_2 \ket{00} - \alpha_3 \ket{01} + \alpha_0 \ket{10} - \alpha_1 \ket{11} \right) \sigma_y \ket{\varphi_1}  \sigma_y\ket{\varphi_2} + \left( \alpha_3 \ket{00} - \alpha_2 \ket{01} - \alpha_1 \ket{10} + \alpha_0 \ket{11} \right) \sigma_z \ket{\varphi_1} \sigma_z \ket{\varphi_2} \right] \\
= & \frac{1}{2} \left[ \ket{00}\left( \alpha_0 II + \alpha_1 \sigma_x \sigma_x + \alpha_2 \sigma_y \sigma_y + \alpha_3 \sigma_z \sigma_z \right) \ket{\varphi_1}\ket{\varphi_2} + 
\ket{01}\left( \alpha_1 II + \alpha_0 \sigma_x \sigma_x - \alpha_3 \sigma_y \sigma_y - \alpha_2 \sigma_z \sigma_z \right) \ket{\varphi_1}\ket{\varphi_2}  + \right. \nonumber \\
&~~\left. \ket{10}\left( \alpha_2 II - \alpha_3 \sigma_x \sigma_x + \alpha_0 \sigma_y \sigma_y - \alpha_1 \sigma_z \sigma_z \right) \ket{\varphi_1}\ket{\varphi_2} + 
\ket{11}\left( \alpha_3 II - \alpha_2 \sigma_x \sigma_x - \alpha_1 \sigma_y \sigma_y + \alpha_0 \sigma_z \sigma_z \right) \ket{\varphi_1}\ket{\varphi_2} \right]
\end{align}
Next, according to the measurement results (in the computational basis) of the two control qubits, different Pauli gates are applied to the two target qubits:
\begin{align}
I\otimes I \left(\alpha_0 II +\alpha_1 \sigma_x \sigma_x + \alpha_2 \sigma_y \sigma_y + \alpha_3 \sigma_z \sigma_z\right) \to \alpha_0 II +\alpha_1 \sigma_x \sigma_x + \alpha_2 \sigma_y \sigma_y + \alpha_3 \sigma_z \sigma_z \\
(\sigma_x \otimes \sigma_x) \left(\alpha_1 II + \alpha_0 \sigma_x \sigma_x - \alpha_3 \sigma_y \sigma_y - \alpha_2 \sigma_z \sigma_z\right) \to  \alpha_0 II +\alpha_1 \sigma_x \sigma_x + \alpha_2 \sigma_y \sigma_y + \alpha_3 \sigma_z \sigma_z \\
(\sigma_y \otimes \sigma_y) \left(\alpha_2 II - \alpha_3 \sigma_x \sigma_x + \alpha_0 \sigma_y \sigma_y - \alpha_1 \sigma_z \sigma_z\right) \to  \alpha_0 II +\alpha_1 \sigma_x \sigma_x + \alpha_2 \sigma_y \sigma_y + \alpha_3 \sigma_z \sigma_z \\
(\sigma_z \otimes \sigma_z) \left(\alpha_3 II - \alpha_2 \sigma_x \sigma_x - \alpha_1 \sigma_y \sigma_y + \alpha_0 \sigma_z \sigma_z\right) \to \alpha_0 II +\alpha_1 \sigma_x \sigma_x + \alpha_2 \sigma_y \sigma_y + \alpha_3 \sigma_z \sigma_z
\end{align}
Therefore, the linear-combination operation $\alpha_0 II + \alpha_1 \sigma_x \sigma_x + \alpha_2 \sigma_y \sigma_y + \alpha_3 \sigma_z \sigma_z$ can be implemented deterministically. Together with $P_1$, $P_2$, $Q_1$ and $Q_2$, the linear-combination operation $U =\sum_{i=0}^{3}{\alpha_i \left(P_1\sigma_i Q_1\right) \otimes \left(P_2\sigma_i Q_2\right)}$ can be implemented deterministically. 

The two control qubits can be replaced by a single ququard and then $U_{\text{LC}}$ can be implemented as a single-ququard operation. It is also worth-noting that there is no limits to the size of each linear term in principle and thus a deterministic linear-combination of four larger gates can be implemented similarly. 

\section{Optical Linear-combination implementation for universal two-qubit unitary gates}

\subsection{Further discussion of optical linear-combination scheme}
The proposed optical scheme for implementing linear combination adopted the Hilbert space extension approach. On our chip, only spatial degree of freedom of each photon is used. Other degrees of freedom can also be used in the optical scheme for encoding high-dimensional information, such as polarization, time-bin and orbital angular momentum. 

The optical linear-combination scheme is more resource-efficient than other existing optical schemes for implementing universal two-qubit unitaries. We compare it with two main schemes for universal linear optical quantum computing: the KLM scheme and MBQC scheme. In standard quantum circuit model, the universal two-qubit quantum circuit requires at least three CNOT and eight single-qubit gates~\cite{vidal2004universal}, as shown in Fig.~\ref{fig:universalthreecnot}(A). In linear optical systems, a CNOT or other equivalent two-qubit-entangling logic gates~\cite{o2009photonic} can only be implemented probabilistically, if requiring no optical nonlinearity~\cite{knill2001scheme}. KLM scheme implements a CNOT gate probabilistically on coincidence detection of ancillary photons at two single photon detectors, and the success probability is $1/16$~\cite{knill2001scheme}. There are also some other protocols proposed for implementing CNOT gate with no optical nonlinearity, such as the protocols proposed by T. B. Pittman \textit{et al.} \cite{pittman2001probabilistic} and by Hofmann \& Takeuchi~\cite{hofmann2002quantum} and by T. C. Ralph \textit{et al.}~\cite{ralph2002linear}. Pittman's protocol utilizes four photons and implements a CNOT with the success probability of 1/4. Ralph's and Hofmann \& Takeuchi's protocol implements a CNOT by using only two photons with the success probability of 1/9. However, post-selection of the coincidence events was required, and thus it cannot be used for implementing cascaded CNOT gates. Therefore, the most resource-efficient way for implementing a quantum circuit shown in Fig.~\ref{fig:universalthreecnot}(A) is to use two Pittman CNOT gates and one post-selected CNOT gate, which requires 6 photons totally and has the overall success probability of $1/144$.

MBQC scheme differs from standard quantum circuit model~\cite{raussendorf2001one,barz2015quantum,kiesel2005experimental,prevedel2007high,barz2012demonstration}, which implements quantum computation by performing a sequence of single-qubit measurements onto the prepared cluster state. MBQC is less demanding of resources than the standard circuit model~\cite{browne2005resource}. However, MBQC requires at least a 6-photon cluster state (Fig.~\ref{fig:universalthreecnot}(B)) to implement the computation equivalent to the universal two-qubit circuit shown in Fig.~\ref{fig:universalthreecnot}(A). It needs more photons to create such a state, and has low success probability as $1/128$, where we assume the required cluster state is created via Type-I fusion~\cite{browne2005resource} with success probability of $1/2$ each.

\begin{figure}[htbp!] 
\centering    
\includegraphics[width=0.6\textwidth]{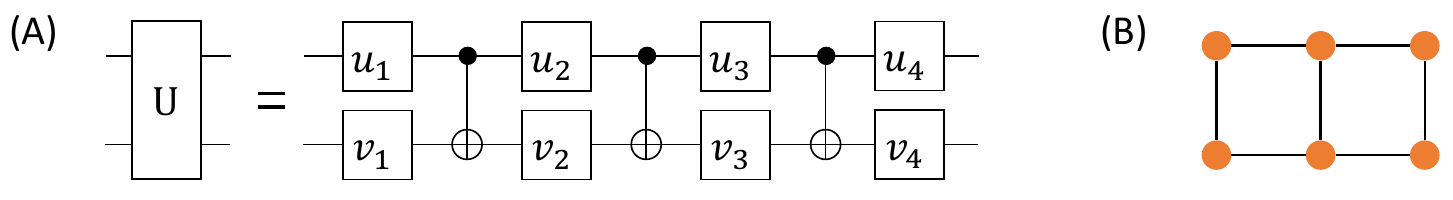}
\caption[Universal two-qubit quantum computation circuit consisting of three consecutive CNOT gates]{(A) Quantum circuit for universal two-qubit quantum operation $U \in \text{SU(4)}$, requiring three consecutive CNOT gates and eight single-qubit gates $u_i$ and $v_i$ ($i=1,\cdots,4$)~\cite{vidal2004universal}. (B) A six-photon cluster state required for implementing QIP that utilizes at least three two-qubit entangling operations in MBQC scheme.}
\label{fig:universalthreecnot}
\end{figure}

\subsection{Advanced photonic linear-combination circuit design for higher success probability}
Our current chip implements a two-qubit unitary operation with success probability of 1/64. It can be further improved to 1/4, if we (i) can separate signal and idler photons with certainty; (ii) use an advanced circuit design as shown in Fig.~\ref{fig:fullLinearCombCircuit}. The new circuit design will utilize the unused optical ports (in our current chip), and the circuit succeeds when the coincidence detections happen between the ports 1 and 2 or $1'$ and $2'$ or $1''$ and $2''$ or $1'''$ and $2'''$.

\begin{figure}[htbp!] 
\centering    
\includegraphics[width=0.6\textwidth]{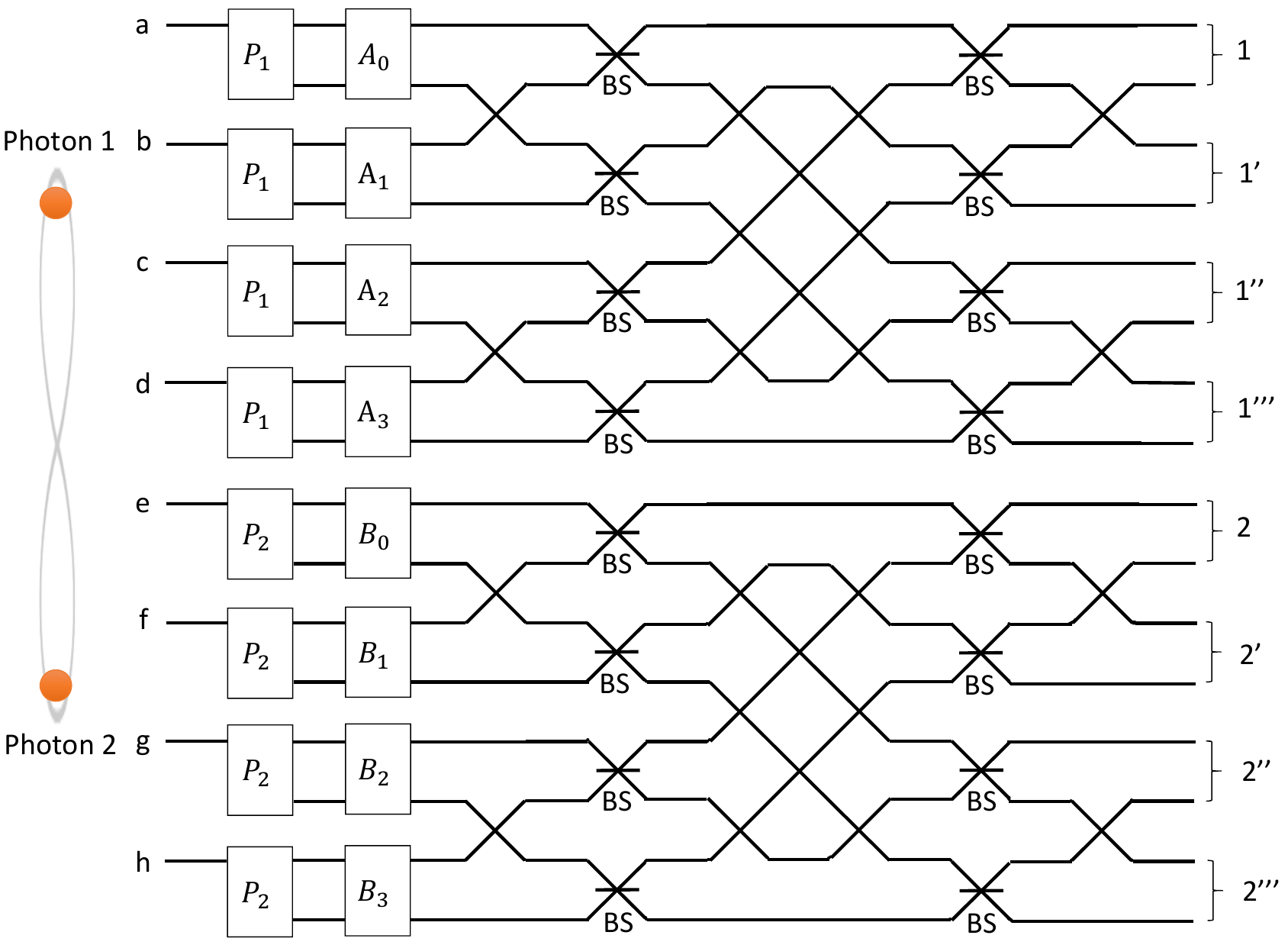}
\caption[]{Schematic of an advanced circuit design for implementing $\text{SU}(4)$. The entangled ququards creation part and tomography measurement part are omitted. Assume photon in top path represents the basis state $\ket{0}$ and the bottom represents $\ket{1}$, $P_1$ and $P_2$ convert the single-qubit state $\ket{0}$ into two arbitrary single-qubit states $\ket{\varphi_1}$ and $\ket{\varphi_2}$ respectively. $A_0$ - $A_3$, $B_0$ - $B_3$ are single-qubit operations. The linear-combination operation succeeds when the coincidence events detected where two photons exit at ports $1$, $2$ (or $1'$, $2'$; $1''$, $2''$; $1'''$, $2'''$) respectively.}
\label{fig:fullLinearCombCircuit}
\end{figure}

\section{Details of Device and Experimental Setup}
\subsection{{Device}}
Our silicon chip was fabricated using a 248 nm lithography at the IME foundry. The starting substrate is silicon-on-insulator (SOI) substrate with 2 $\mu \text{m}$ buried oxide (BOX) and 220 nm top crystalline Si on BOX. The 500-nm-wide silicon waveguides were patterned in the 220-nm top layer of silicon. Resistive heaters working as thermo-optic phase shifters, 180 $\mu \text{m}$ long and 2.5 $\mu \text{m}$ wide, are then patterned on a 120 nm TiN metal layer on top of the waveguide layer. The effective footprint of the device is approximately $7.1 \times 1.9 ~\text{mm}^2$. 

The propagation loss for channel waveguide is 3 dB/cm in our device. The multimode interferometers (MMI) in the device has around 0.3 dB insertion loss. The thermo-optic phase shifters have ignorable loss compared to the MMIs. The electric power required for $2\pi$ phase shift for one individual thermo-optic phase shifter is around 57 mW. All the thermo-optic phase shifters each have about 800 $\Omega$ resistance and the applied current for $2 \pi$ phase shift is less than 10 mA.

\subsection{{Setup}}
The schematic diagram of the whole setup was shown in Fig. 1(D) in the main text. Bright light with the wavelength of 1550.8 nm is collected from a tunable laser (Yenista Optics Tunics-T100S-HP) and further amplified using an EDFA (Pritel), up to 300 mW. The amplified spontaneous emission (ASE) noise is suppressed using a DWDM module filter (Opneti). The bright light is then injected into the silicon chip via a 48-channel V-Groove fiber array (OZ-optics) with 127 $\mu \text{m}$ spacing and 10-degree angle polished from top to bottom. An in-line polarization controller (FiberPro) is used to ensure that TE-polarized light is launched into the chip through TE grating couplers fabricated on the chip. 

At the end of the optical output of the chip, two off-chip DWDM module filters, having a 200 GHz channel space and 1 nm 0.5 dB bandwidth, are used to separate the signal and idler photons. We chose the channels that are equally 4-channels away from the channel used for the pump light for selecting the signal and idler photons respectively: $\lambda_p - \lambda_s = \lambda_i - \lambda_p = 6.6$ nm. We only picked up the signal photons ($\lambda_s = 1544.2$ nm) through one DWDM filter ($\text{DWDM}_1$) and the idler photons ($\lambda_i = 1557.4$ nm) through the other ($\text{DWDM}_2$), realizing the post-selected measurements. Note that we can also pick up the idler photons from ($\text{DWDM}_1$) and the signal photons from ($\text{DWDM}_2$) to realize post-selection, which can in theory double the coincidence counts but require two extra single-photon detectors. 

For the signal photon channel and the idler photon channel, the insertion losses over full passband are 1.89 dB and 2.88 dB respectively. Both of the DWDM filters have $\ge$ 45 dB isolation over full passband for non-adjacent channels. Together with the on-chip pump filters, the pump injected to the chip are effectively isolated from the photon detections. The coupling loss of our device (including both input and output) is 13 dB, which is estimated by injecting pump light through V-Groove fiber array to a long straight waveguide surrounding our device on the silicon chip. Chip expansion happens when the phase shifters are heated up, and thus introduces instability of optical coupling. To overcome this issue, we used a Peltier cell controlled by a proportional integrative derivative (PID) controller to keep the device temperature constant actively.

The photons are detected by two fiber-coupled superconducting nanowire single-photon detectors (SNSPDs) mounted in a closed cycle refrigerator. Two in-line polarization controllers are used to optimize the polarizations of photons before going into SNSPDs, which maximizes the detection efficiency up to 50\%. Two-photon coincidences are recorded by using a time-interval-analyzer (PicoHarp 300) in a 450 ps integration window, which was chosen according to the jitter-time of SNSPDs. During the experiments, the dark count for single photon detection is 800-1500 per second, but is quite trivial for the two-photon coincidence events. Upon using 300 mW pump power, we obtain around 100 coincidence counts per second during the experiments.

\section{{Characterization of the device}}
\subsection{Electrical characteristics of the phase shifters}
Characterization is critical for integrated quantum photonic devices. In our device, the main task is to calibrate the electric power-optics responses of the phase shifters, i.e., find the corresponding phase of a phase shifter for each applied voltage or current. All the thermal-optic phase shifters are controlled by 62-channel current output equipment (UEI AO current boards), which allows to set the required current to each channel via serial communication from a classical computer. 

We first measured the resistance value of each phase shifter by scanning its I-V curve. The I-V curve of an example phase shifter is shown in Fig.~\ref{fig:IVCurve}, showing perfect linearity. 
\begin{figure*}[htbp!] 
\centering    
\includegraphics[scale=0.6]{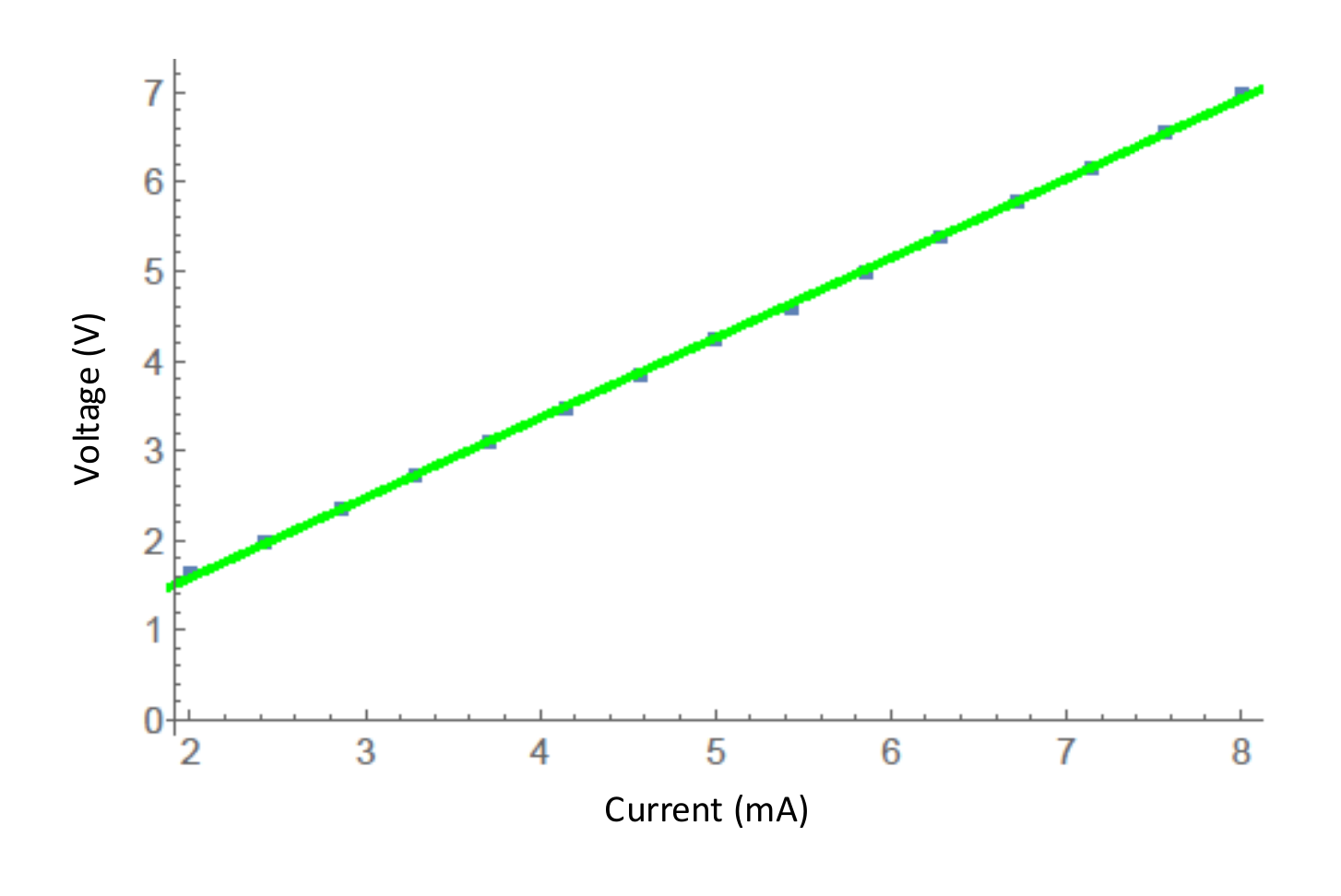}
\caption[Measured I-V curve of an example thermal phase shifter]{The measured I-V curve of an example thermal-optic phase shifter. The horizontal axis represents the current applied to the phase shifter and the vertical axis represents the measured voltage.}
\label{fig:IVCurve}
\end{figure*}
The obtained I-V curve can be simply fitted into a linear equation as 
\begin{align}
V= R\cdot I + \delta V
\end{align}
where $R$ represents the resistance value of the phase shifter and $\delta V$ represents the offset of the voltage of the current output boards with 0 mA current setting. The fitted resistances of all the phase shifters in our device are around 800 $\Omega$, except that the four phase shifters used in the pump filters have resistance values of around 580 $\Omega$.

Thermal-optic phase shifter has a linear phase-(electrical)power relationship and further a nonlinear phase-current relationship, considering the fact that elect-power of a phase shifter is given by
\begin{align}
P = I^2R.
\end{align} 
Therefore, the phase-current relationship of a phase shifter can be obtained as
\begin{align}
\theta(I) = \phi_1 I^2 + \phi_0 \label{eq:fittingheater}
\end{align}
where $I$ is the current applied to the phase shifter and $\theta(I)$ is the resulting phase shift. $\phi_1$ and $\phi_0$ are real numbers associated with the response of a particular phase shifter. 

All the 62 thermal-optic phase shifters are divided into two kinds: independent phase shifters and cascaded phase shifters. Here we say an independent phase shifter is the one occupying a MZI whose input and output can be directly accessed by external laser and optical power-meter. For the independent phase shifter, we assume the associated MMIs have good splitting ratio close to 50:50. In fact, we measured that such an independent MZI in our device can achieve an extinction ratio up to 30 dB~\cite{wilkes2016high}.

\subsection{Calibrating independent phase shifters}
An example independent phase shifter is shown in Fig.~\ref{fig:IndependentPhaseShifter}(A), where we can directly inject bright light (1550 nm, 20 mW) into one input port (for example, In1) of the MZI and measure the intensity at one output port (Out1) as a function of the current applied to the heater. The applied current increases linearly from 0 mA to 9 mA with the step size of 0.05 mA. The obtained classical interference fringe of one example independent phase shifter is shown in Fig.~\ref{fig:IndependentPhaseShifter}(B). 

Using nonlinear fitting approach (such as Mathematica built-in function: NonlinearModelFit), we were able to fit the experimental data with Equation~\eqref{eq:fittingheater}, obtaining that $\phi_1 = 0.1123$ and $\phi_0 = 0.3814$. In our device, the phase shifters used in the four filters, in setting the laser intensities between four SFWM sources and in the stage of performing measurement are independent phase shifters that can be calibrated directly. The phase shifters in setting phases between four SFWM sources and in the stage of realizing linear-combination can also be calibrated as the independent phase shifter once the cascaded phase shifters in the stages of preparing single-qubit states and implementing operations were calibrated, which we will discuss in the following section.

\begin{figure}[htbp!] 
\centering    
\includegraphics[scale=0.7]{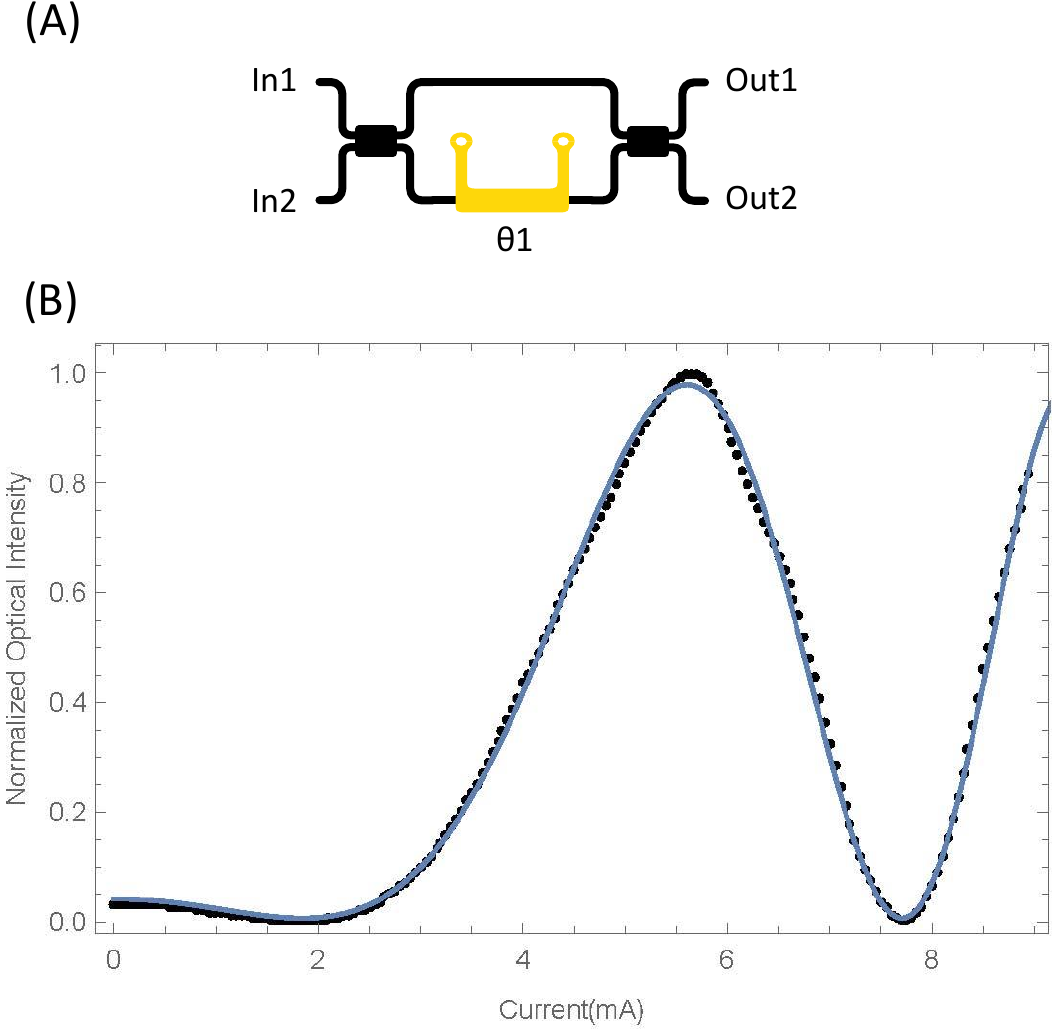}
\caption{(A) Schematic diagram of an independent phase shifter occupying in an MZI. (B) The current optical-intensity fringe curve of an example independent phase shifter. The black points represent the experimental data and the blue curve is obtained through the fitting function, showing a good fitness.}
\label{fig:IndependentPhaseShifter}
\end{figure}

\subsection{Calibrating cascaded phase shifters}
The cascaded phase shifters illustrated in Fig.~\ref{fig:ConsecutivePS} the five phase shifters located in an array---used in the stages of preparing single-qubit state and implementing gates, as shown in the Fig. 1(D) in the main text. Here we present a schematic diagram of such cascaded phase shifters in Fig.~\ref{fig:ConsecutivePS}. There are eight cascaded phase shifter arrays in our device. These phase shifters cannot be simply calibrated like the independent ones, since we can only access the input and output ports of the whole array. 

\begin{figure}[htbp!] 
\centering    
\includegraphics[scale=0.6]{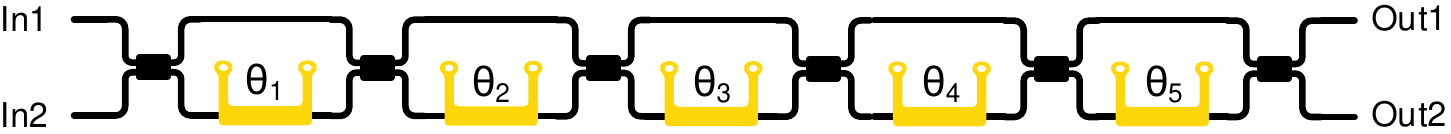}
\caption{Schematic of five cascaded phase shifters in an array.}
\label{fig:ConsecutivePS}
\end{figure}

To calibrate these phase shifters, we scanned a 5-dimensional (5D) fringe for each array and performed multiple parameters fitting. In the 5D scan, we scan same current range for each of the five phase shifters and record the experimentally obtained output power for every current configuration. This 5D scan fitting approach works well for our device, but it may be challenging for a larger number of cascaded phase shifters since it requires exponentially-increasing amount of data.

Specifically, we injected the classical light into one input port (In1) and measured the output intensity at one output port (Out1). For a current configuration of the five phase shifters $\{I_1, I_2, I_3, I_4, I_5 \}$, we measured the corresponding output intensity. A least-square minimization algorithm was used to fit the measured optical intensities and theoretical values, and thus the fitted parameters associated with the responses of the phase shifters can be estimated. The theoretical transfer matrix of the cascaded phase shifters is given by 
\begin{align}
&U_{\text{Array}}[\eta_0, \eta_1, \eta_2, \eta_3, \eta_4, \eta_5, \theta_1, \theta_2, \theta_3, \theta_4, \theta_5] \nonumber \\ 
& = \text{BS}[\eta_5]\text{PS}[\theta_5]\text{BS}[\eta_4]\text{PS}[\theta_4]\text{BS}[\eta_3]\text{PS}[\theta_3]\text{BS}[\eta_2]\text{PS}[\theta_2]\text{BS}[\eta_1]\text{PS}[\theta_1]\text{BS}[\eta_0] 
\end{align}
where BS and PS define the transfer matrices of MMI beam-splitter and phase shifter as follows:
\begin{align}
\text{BS}[\eta]&:=\begin{pmatrix} \sqrt{\eta} & i\sqrt{1-\eta}\\ i\sqrt{1-\eta} & \sqrt{\eta} \end{pmatrix}, \\
\text{PS}[\theta]&:=\begin{pmatrix} 1 & 0\\0  & e^{i\theta} \end{pmatrix}.
\end{align} 
Note that here $\eta$ represents the splitting ratio of each MMI, which is close to 50:50 and independent of applied current. $\theta_i$ represents the phase setting of the corresponding phase shifter, which is given by 
\begin{align}
\theta_i = \phi_i(I_i^2 R_i) + \delta\theta_i
\label{eq:phasecurr}
\end{align}
where $I_i$ is the current applied to the heater $i$ with resistance value $R_i$, and $\delta \theta_i$ is the phase offset when zero-current applied. 

The objective function of minimization is thus obtained as 
\begin{align}
\text{Min}\left(\sum_m{\left(O_m - A\cdot \left|\bra{0}U_\text{Array}[\eta_0, \eta_1, \eta_2, \eta_3, \eta_4, \eta_5, \theta_1, \theta_2, \theta_3, \theta_4, \theta_5]\ket{0}\right|^2\right)^2}\right)
\end{align}
where $O_m$ is normalized experimentally obtained optical intensity and $\theta_i$ is replaced by using Equation~\eqref{eq:phasecurr}. The minimization process can be implemented using the Mathematica built-in function ``FindMinimun'' with proper initial guess. After the fitting process, we obtain the fitted results for the parameters $\eta_0$, $\eta_1$, $\eta_2$, $\eta_3$, $\eta_4$, $\eta_5$, $\phi_1$, $\delta \theta_1$ $\phi_2$, $\delta \theta_2$ $\phi_3$, $\delta \theta_3$ $\phi_4$, $\delta \theta_4$ $\phi_5$, $\delta \theta_5$. These 16 parameters are able to describe the behaviors of the response of the calibrated cascaded phase shifter array for each current configuration, making each phase shifter in the array fully reconfigurable.

\subsection{Calibrating on-chip filters}
In our device, the on-chip filters are used to remove the redundant pump light out from the photon sources. The on-chip pump filter is implemented by using an imbalanced MZI, as shown in Fig~\ref{fig:PumpFilter}. 

\begin{figure}[htbp!] 
\centering    
\includegraphics[scale=0.6]{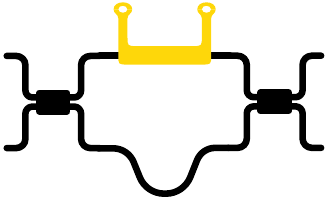}
\caption[Schematic of the on-chip pump filter]{Schematic of the on-chip pump filter. On the actual silicon chip, the arm where the phase shifter locates is 54 $\mu \text{m}$ longer than the other arm of MZI.}
\label{fig:PumpFilter}
\end{figure}

With a proper current applied, the imbalanced MZI has different transmission responses for different wavelengths of light, as shown in Fig.~\ref{fig:PumpFilterSpectra}. It shows that the pump filter has very low transmission rate for the light with wavelength around 1550.8 nm but high transmission rate for the light with wavelengths around 1544.2 nm and 1557.4 nm---which corresponds to the signal and idler photons respectively. The extinction ratio of the pump filter is up to $\sim28$ dB, which filters the pump light after the photon sources quite effectively. In the experiments, we used a laser pump of up to 300 mW (24.77 dBm) for each SFWM photon source. After the on-chip pump filter, there is $\sim0.475$ mW (-3.223 dBm) pump light passing through the functional area of the device (states and operations configuration and measurement, marked 2 to 5 in Figure 1 D of the main text). This is a very small intensity of pump which only generates trivial SFWM effect during the waveguides in the functional area, adding negligible counts to the signal that are suppressed through our measurement of the coincidence counts between signal and idler photons generated from the same source---the signal photon and the idler photon of the same pair pass through the top half and bottom half of the device separately. But the photon pairs generated from remnant of the pump will always pass through the same half of the device, top or bottom, which is impossible to cause a coincidence count in our detection setup.

\begin{figure*}[htbp!] 
\centering    
\includegraphics[width=0.8\textwidth]{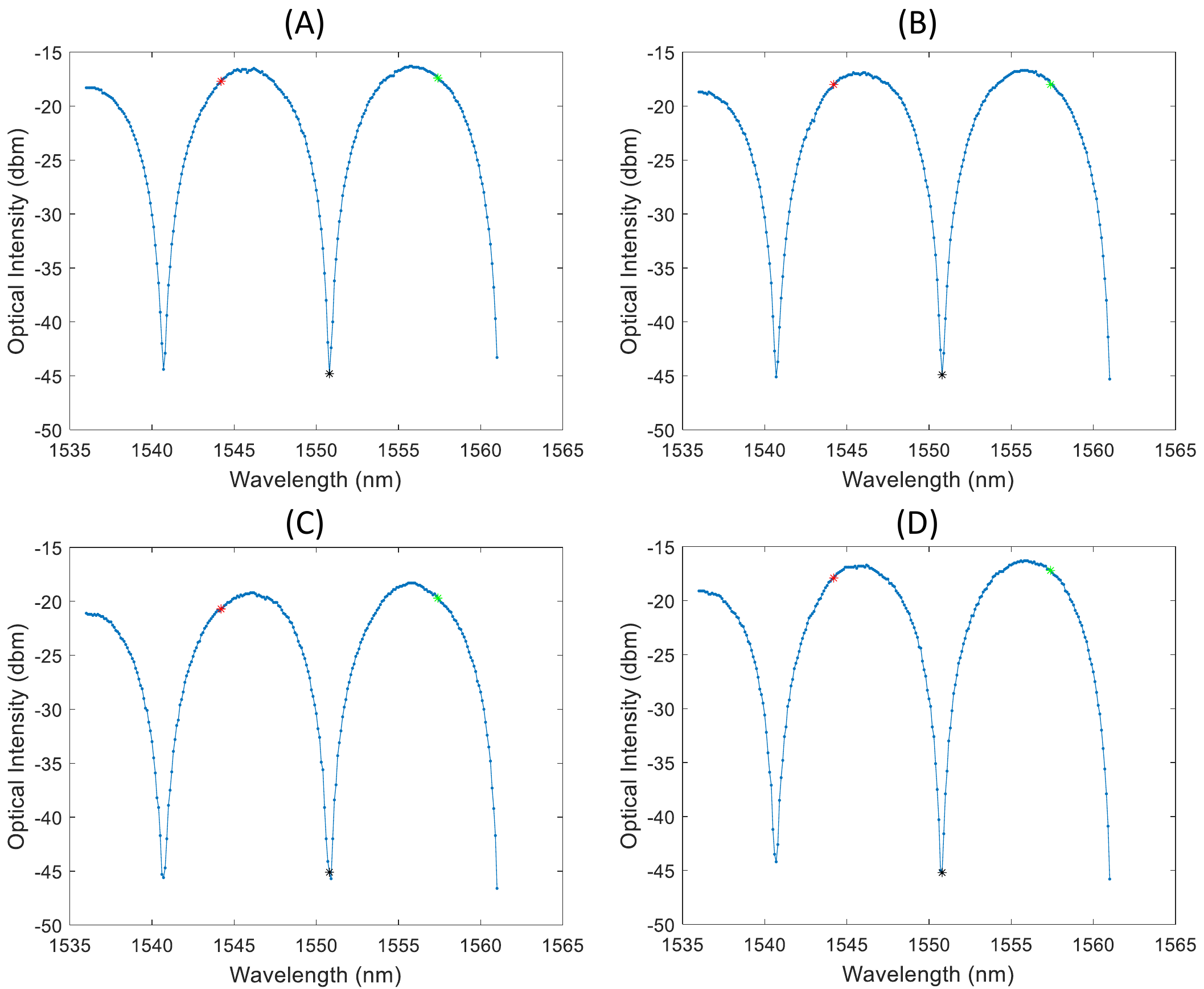}
\caption[The transmission response of the on-chip pump filters for different wavelengths]{Transmission responses of the four on-chip pump filters with wavelengths. The pump light is injected into the top input port and is measured at the top output port. {The red, green and black stars in figures mark the signal, idler and pump wavelengths respectively.}
 }
\label{fig:PumpFilterSpectra}
\end{figure*}

\section{{Further details of experimental quantum processing results}}
\subsection{Implementing two-qubit quantum logic gates}
Here we first present details for implementing some specific quantum gates: CNOT, CZ, CH, SWAP, iSWAP and $\sqrt{\text{SWAP}}$. The linear-combination decompositions of CNOT, CZ and CH are listed as follows:
\begin{align}
\text{CNOT} = \frac{1}{\sqrt{2}}\begin{pmatrix} 1 & 0\\ 0 & i \end{pmatrix} \otimes \frac{1}{\sqrt{2}}\begin{pmatrix} 1 &-i\\ -i & 1 \end{pmatrix} + \frac{1}{\sqrt{2}}\begin{pmatrix} 1 & 0\\ 0 &-i \end{pmatrix} \otimes \frac{1}{\sqrt{2}}\begin{pmatrix} 1 & i\\ i & 1 \end{pmatrix}
\label{eq:CNOT}
\end{align}
\begin{align}
\text{CZ} = \frac{1}{\sqrt{2}}\begin{pmatrix} 1 & 0\\ 0 & i \end{pmatrix} \otimes \frac{1-i}{\sqrt{2}}\begin{pmatrix} 1 &0\\ 0 & i \end{pmatrix} + \frac{1}{\sqrt{2}}\begin{pmatrix} 1 & 0\\ 0 &-i \end{pmatrix} \otimes \frac{1+i}{\sqrt{2}}\begin{pmatrix} 1 & 0\\ 0 & -i \end{pmatrix}
\label{eq:CZ}
\end{align}
\begin{align}
\text{CH} = \frac{1}{\sqrt{2}}\begin{pmatrix} 1 & 0\\ 0 & i \end{pmatrix} \otimes \frac{1}{\sqrt{2}}\begin{pmatrix} \sqrt{2}-i &-i\\ -i & \sqrt{2}+i \end{pmatrix} + \frac{1}{\sqrt{2}}\begin{pmatrix} 1 & 0\\ 0 &-i \end{pmatrix} \otimes \frac{1}{\sqrt{2}}\begin{pmatrix} \sqrt{2}+i & i\\ i & \sqrt{2}-i \end{pmatrix}
\label{eq:CH}
\end{align}
To implement these gates in our device, we first configured the device to to create the following maximally path-entangled state:
\begin{align}
\frac{1}{\sqrt{2}}\ket{1}_a\ket{1}_e + \frac{1}{\sqrt{2}}\ket{1}_b\ket{1}_f,
\end{align}
and then programmed the local single-qubit operations on path $a$, $b$, $e$ and $f$, following Equations~\eqref{eq:CNOT}, \eqref{eq:CZ} and \eqref{eq:CH}. The initial input states for computation can be arbitrarily configured through the state preparation stages (including the first two phase shifters in each cascaded phase shifter array). The obtained process fidelities for these three gates are 98.85$\pm$0.06\%, 96.90$\pm$0.17\% and 97.57$\pm$0.07\% respectively.

It is the first time for implementing SWAP, iSWAP and $\sqrt{\text{SWAP}}$ gates in integrated photonics chips. In standard quantum circuit model, a SWAP gate is implemented by a sequence of three CNOT gates where the middle one reverses the control and target qubits. An iSWAP gate can be implemented by a unitary gate $(I\otimes I - i\sigma_z\otimes \sigma_z)$ followed by a SWAP gate. $\sqrt{\text{SWAP}}$ gate performs half-way of SWAP, which allows for universal quantum computation together with single-qubit gates. The corresponding transfer matrices of these three gates are as follows.
\begin{align}
\text{SWAP} = \begin{pmatrix} 1&0&0&0 \\0&0&1&0 \\ 0&1&0&0 \\ 0&0&0&1  \end{pmatrix},~
\text{iSWAP} = \begin{pmatrix} 1&0&0&0 \\0&0&i&0 \\ 0&i&0&0 \\ 0&0&0&1  \end{pmatrix},~
\sqrt{\text{SWAP}}=\begin{pmatrix} 1&0&0&0 \\0&\frac{1+i}{2}&\frac{1-i}{2}&0 \\ 0&\frac{1-i}{2}&\frac{1+i}{2}&0 \\ 0&0&0&1  \end{pmatrix}
\end{align}
The linear-combining decompositions of the three gates are  
\begin{align}
\text{SWAP} &= \frac{1}{2}\left( I\otimes I + \sigma_x\otimes \sigma_x + \sigma_y\otimes \sigma_y + \sigma_z\otimes \sigma_z  \right) \\
\text{iSWAP} &= \frac{1}{2}\left( I\otimes I + i\sigma_x\otimes \sigma_x +i \sigma_y\otimes \sigma_y + \sigma_z\otimes \sigma_z  \right) \\
\sqrt{\text{SWAP}}&=(0.75+0.25i) I\otimes I + (0.25-0.25i)\sigma_x\otimes \sigma_x +(0.25-0.25i) \sigma_y\otimes \sigma_y +(0.25-0.25i) \sigma_z\otimes \sigma_z
\end{align}
all of which are simple linear-combinations of tensor products of Pauli gates and identity gates. 

For implementing SWAP and iSWAP, balanced pumps were split between the four photon sources with zero relative phase (SWAP) or $\pi/2$ relative phase (iSWAP), creating the following maximally path-entangled states:
\begin{align}
\text{SWAP}: ~\frac{1}{2}\ket{1}_a\ket{1}_e + \frac{1}{2}\ket{1}_b\ket{1}_f + \frac{1}{2}\ket{1}_c\ket{1}_g + \frac{1}{2}\ket{1}_d\ket{1}_h \\
\text{iSWAP}: ~\frac{1}{2}\ket{1}_a\ket{1}_e + \frac{i}{2}\ket{1}_b\ket{1}_f + \frac{i}{2}\ket{1}_c\ket{1}_g + \frac{1}{2}\ket{1}_d\ket{1}_h 
\end{align} 
For implementing $\sqrt{\text{SWAP}}$, the path-entangled state will be required as
\begin{align}
(0.75+0.25i)\ket{1}_a\ket{1}_e + (0.25-0.25i)\ket{1}_b\ket{1}_f + (0.25-0.25i)\ket{1}_c\ket{1}_g + (0.25-0.25i)\ket{1}_d\ket{1}_h
\end{align}
and thus requires unbalanced pumps split between the four photon sources.  
The obtained process fidelities of SWAP, iSWAP and $\sqrt{\text{SWAP}}$ are 95.33$\pm$0.24\%, 94.45$\pm$0.27\%, 92.41$\pm$0.33\% respectively. The reconstructed process matrices of these implemented gates are shown in Fig.~\ref{fig:ProcessMatricesOfGates}. 

For showcasing performance of the chip, we also implemented many other two-qubit quantum logic gates. These gates include four different kinds of gates, which are in the forms of $A_1\otimes B_1$, Controlled-$U$, $\alpha A_1\otimes A_2 + \beta B_1 \otimes B_2$ and $\alpha A_1 \otimes A_2 + \beta B_1 \otimes B_2 + \gamma C_1 \otimes C_2 + \delta D_1 \otimes D_2$ respectively. Here $A_1$ ($A_2$), $B_1$ ($B_2$), $C_1$ ($C_2$), $D_1$ ($D_2$), $U$, $\alpha$, $\beta$, $\gamma$ and $\delta$ are randomly chosen for each instance of the tested gates. These gates utilize a quarter, a half or full capability of the chip, demonstrating the full performance of the chip in various cases. Together with the specific gates mentioned above, we implemented 98 two-qubit quantum logic gates, performed quantum process tomography and reconstructed the corresponding process matrices for each instance, achieving the process fidelity of 93.15$\pm$4.53\% in average. During the experiments, the typical integration time for each count of process tomography is 10 seconds, and the total cost time for performing full quantum process tomography for one tested gate is more than 100 minutes.

\begin{figure*}[htbp!] 
\centering    
\includegraphics[width=1.0\textwidth]{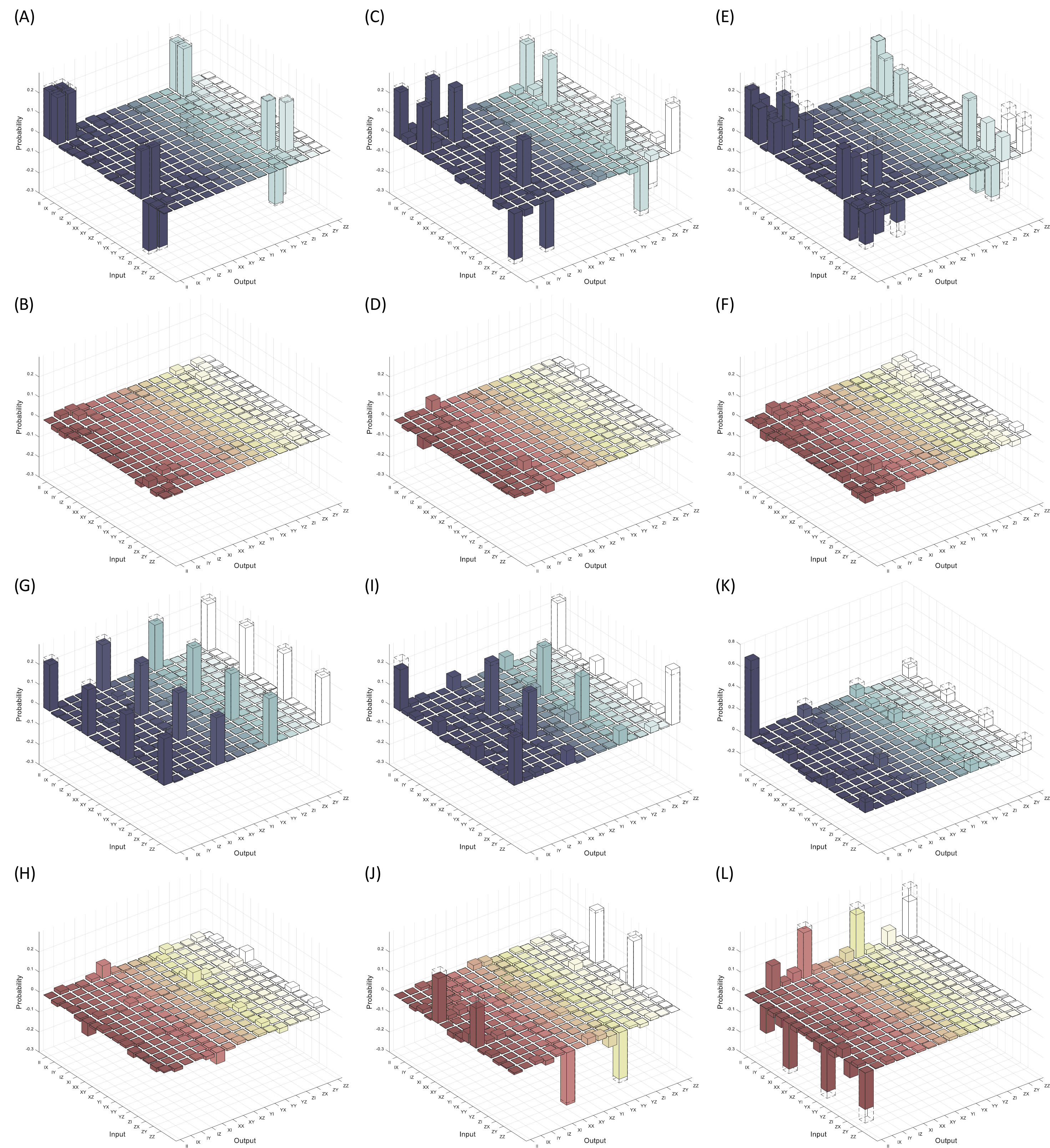}
\caption{Experimentally reconstructed process matrices (solid colour) of implemented example quantum logic gates, with theoretical prediction represented by overlaid empty frames. (A, B) The real and imaginary parts of the process matrix of CNOT. (C, D) The real and imaginary parts of the process matrix of CZ. (E, F) The real and imaginary parts of the process matrix of CH. (G, H) The real and imaginary parts of the process matrix of SWAP. (I, J) The real and imaginary parts of the process matrix of iSWAP. (K, L) The real and imaginary parts of the process matrix of $\sqrt{\text{SWAP}}$.}
\label{fig:ProcessMatricesOfGates}
\end{figure*}

\subsection{{Quantum process tomography}}
Quantum process tomography (QPT) is a well-known approach for experimentally characterizing a quantum operation, which is a complex and resource-intensive procedure providing a complete characterization of the implemented gates in our chip. To perform QPT for each implemented two-qubit gate, we configured 16 separable, linearly independent states $\{\rho_i^{(in)} \} = \{\ket{\psi_i}\bra{\psi_i}\}$ as the inputs of the implemented gate, where $\ket{\psi_i}=\ket{\nu_1}\otimes \ket{\nu_2}$ and $\{\ket{\nu_1},\ket{\nu_2}\} = \{ \ket{0}, \ket{1}, \ket{+} = \frac{1}{\sqrt{2}}(\ket{0}+\ket{1}), \ket{+i} = \frac{1}{\sqrt{2}}(\ket{0}+i\ket{1})\}$. For each input, we measured the output state in a set of 16 basis projections
\begin{align}
\{ &\ket{00}\bra{00} , \ket{01}\bra{01},\ket{0,+}\bra{0,+},\ket{0,+i}\bra{0,+i}, \nonumber \\
&\ket{10}\bra{10} , \ket{11}\bra{11},\ket{1,+}\bra{1,+},\ket{1,+i}\bra{1,+i}, \nonumber  \\ 
&\ket{+,0}\bra{+,0} , \ket{+,1}\bra{+,1},\ket{+,+}\bra{+,+},\ket{+,+i}\bra{+,+i},\nonumber \\
&\ket{+i,0}\bra{+i,0} , \ket{+i,1}\bra{+i,1},\ket{+i,+}\bra{+i,+},\ket{+i,+i}\bra{+i,+i}\},
\end{align} 
and the density matrix of the output state can be experimentally reconstructed via quantum state tomography. On the other hand, the output state can be theoretically obtained as below:
\begin{align}
\textstyle \rho_i^{(out)}= \sum_{m,n=0}^{d^2-1}{\chi_{mn}\hat{A}_{m}\rho_i^{(in)}\hat{A}_n^{\dag}}
\end{align}
where $\hat{A}_m$ are the Kraus operators and $\chi$ is the corresponding process matrix of the implemented gate. The process matrix $\chi$ was reconstructed by using the maximum likelihood techniques, which completely describes the implemented gate. For each implemented gate, QPT requires 256 (=$16\times 16$) coincidence measurements in total.

The process fidelity was calculated as $F_P=\text{Tr}(\chi_\text{ideal}\chi_\text{exp})$, where $\chi_\text{ideal}$ and $\chi_\text{exp}$ represent the theoretically ideal and experimentally reconstructed process matrices respectively. The error-bars of process fidelities were estimated by a Monte-Carlo approach assuming Poissonian photon statistics.

\subsection{{Demonstrating QAOA}}
Here we present more details of applying QAOA for the three example CSPs shown in the main text.
The objective function of $\text{CSP}_1$ is 
\begin{align}
\textstyle C(z) = \frac{1}{2} + \frac{1}{2}z_1z_2.
\end{align}
According to QAOA process, the two operators with angles $\gamma$ and $\delta$ are obtained as below:
\begin{align}
e^{-i\gamma C} &= e^{-i\gamma [\frac{1}{2}+\frac{1}{2}Z_1Z_2]}=e^{-i\gamma\frac{1}{2}}e^{-i\gamma Z_1Z_2} \\
e^{-i\beta B} &= e^{-i\beta (X_1+X_2)}
\end{align}
where $X$ and $Z$ represent Pauli operators $\sigma_x$ and $\sigma_z$ respectively.
The $\ket{\gamma, \beta}_{\text{CSP}_1}$ state is obtained as 
\begin{align}
\ket{\gamma, \beta}_{\text{CSP}_1}&=e^{-i\beta B} e^{-i\gamma C} H \otimes H \ket{00} \nonumber \\
 &= e^{-i\gamma \frac{1}{2}}e^{-i\beta X_1} e^{-i\beta X_2} e^{-i\gamma \frac{1}{2} Z_1 Z_2} H \otimes H \ket{00}.
\end{align}
It is easy to verify that for $\text{CSP}_1$ two strings of $z$: $\{1,1\}$ and $\{-1,-1\}$ can satisfy the only constraint and make $C(z)=1$.

The objective function of $\text{CSP}_2$ is 
\begin{align}
\textstyle C(z) =\left(\frac{1}{2} + \frac{1}{2}z_1 \right) +\left(\frac{1}{2} + \frac{1}{2}z_2 \right)+  \left( \frac{1}{2} + \frac{1}{2}z_1z_2\right)
\end{align}
and similarly the $\ket{\gamma, \beta}_{\text{CSP}_2}$ state is obtained as 
\begin{align}
\ket{\gamma, \beta}_{\text{CSP}_2}&=e^{-i\beta B} e^{-i\gamma C} H \otimes H \ket{00} \nonumber \\
 &= e^{-i\gamma \frac{3}{2}}e^{-i\beta X_1} e^{-i\beta X_2} e^{-i\gamma \frac{1}{2}\left( Z_1 +Z_2 + Z_1 Z_2  \right)} H \otimes H \ket{00}.
\end{align}
For the strings $z= \{1,1\}, \{1,-1\}, \{-1,1\}, \{-1, -1\}$, the values of $C$ are 3, 1, 1, -1 respectively. Therefore, the string $z=\{1,1\}$ is the target string, and as we showed in the main text, the $p=1$ QAOA outputs $\{z_1,z_2\}=\{1,1\}$ with highest probability. 

The objective function of $\text{CSP}_3$ is 
\begin{align}
\textstyle C(z) =\left(\frac{1}{2} + \frac{1}{2}z_1 \right) +\left(\frac{1}{2} + \frac{1}{2}z_2 \right)+  \left( \frac{1}{2} - \frac{1}{2}z_1z_2\right)
\end{align}
and the $\ket{\gamma, \beta}_{\text{CSP}_3}$ state is obtained as 
\begin{align}
\ket{\gamma, \beta}_{\text{CSP}_3}&=e^{-i\beta B} e^{-i\gamma C} H \otimes H \ket{00} \nonumber \\
 &= e^{-i\gamma \frac{3}{2}}e^{-i\beta X_1} e^{-i\beta X_2} e^{-i\gamma \frac{1}{2}\left( Z_1 +Z_2 - Z_1 Z_2  \right)} H \otimes H \ket{00}.
\end{align}
For the strings $z= \{1,1\}, \{1,-1\}, \{-1,1\}, \{-1, -1\}$, the values of $C$ are 1, 1, 1, -3 respectively. Therefore, there are three strings all can maximize the object function $C$ of ${\text{CSP}_3}$. With proper $\gamma$ and $\beta$, the $p=1$ QAOA outputs the measurement result as $\{z_1,z_2\}=\{1,1\}, \{1,-1\}, \{-1, 1\}$ with approximate probability of 1/3 each. 

Quantum circuits that implement QAOA for the three example CSPs (Fig. 3(A) in the main text) use only one two-qubit entangling gate, and thus they can be implemented using half capability of our chip. The typical integration time for a single measurement of $\ket{\gamma,\beta}$ with each pair of $\{\gamma,\beta\}$ is 15 seconds in our experiment, and the total cost time of experimentally obtaining $\braket{\gamma,\beta|C|\gamma,\beta}$ for all 600 pairs of $\ket{\gamma,\beta}$ is more than 10 hours.

\subsection{{Simulating Szegedy quantum walk}}
The periodicity of SQW can be theoretically analyzed by using spectral decomposition of the single-step evolution operator $U_{\text{sz}}$. $U_{\text{sz}}$ can be decomposed into the following form:
\begin{align}
U_{\text{sz}} = \sum_{i=1}^{N^2}\lambda_i\ket{i}\bra{i},
\end{align} 
where $\lambda_i$ and $\ket{i}$ are eigenvalues and corresponding eigenvectors. A general initial state $\ket{\psi}$ for SQW is given by
\begin{align}
\ket{\psi} = \sum_{i=1}^{N^2} \alpha_i\ket{i},
\end{align}
where $\alpha_i$ satisfy that $\sum_{i=1}^{N^2}{|\alpha_i|^2}=1$. 
The evolution state of a single step of SQW is obtained as
\begin{align}
U_{\text{sz}}\ket{\psi} =\sum_{i=1}^{N^2} \lambda_i \alpha_i\ket{i}, 
\end{align}
and the evolution state of SQW of $t$ steps is determined by
\begin{align}
(U_{\text{sz}})^t\ket{\psi} = \sum_{i=1}^{N^2} \lambda_i^t \alpha_i\ket{i}.
\end{align}

Given that $U_{\text{sz}}$ is unitary, for all eigenvalues $\lambda_i$ of $U_{\text{sz}}$, $|\lambda_i|= 1$. An eigenvalue is said to be a root of unity if there exists a natural number $n$ such that $\lambda^{n} = 1$, where $n$ is the period. The single-step evolution operator $U_{\text{sz}}$ is periodic if and only if all its corresponding eigenvalues are roots of unity with a common period. If $U_{\text{sz}}$ is periodic, its period is given by the lowest common multiple of the periods of its eigenvalues. Using this analysis it was found that SQWs on the given two-node complete graph are periodic for the cases $\alpha =$ $\beta =$ $\frac{1}{4}$, $\frac{1}{2}$,$\frac{3}{4}$ and 1 with periods of 6, 4, 6 and 2 steps respectively. Moreover, SQWs are periodic for the cases $\alpha+\beta=1$ and $\alpha+\beta=\frac{1}{2}$ with periods of 4 and 6 steps respectively.

In our experiments, we implemented SQWs for various cases of weights $\alpha$ and $\beta$ with three different initial states: $\ket{0}\ket{0}$, $\frac{1}{\sqrt{2}}\ket{0}(\ket{0}+\ket{1})$ and $\frac{1}{\sqrt{2}}\ket{0}(\ket{0}+i\ket{1})$. For each case, we realized the first 200 steps of SQW and measured the probability of the walker being at node 1 at each step. We calculated the average fidelities of the obtained probability distributions, as listed in Table.~\ref{my-label}. It is worth noting that SQWs on the example two-node graph do not have perfect periodicity for general symmetric weights, however, the obtained probability distributions may show quasi periodic behaviors for some specific cases. As shown in Figure.~\ref{fig:SQWProbDis}, SQWs for the cases $\alpha=\beta=0.43, 0.45, 0.47$ show quasi-periodicities. This could be useful for designing SQW-based quantum algorithms.

\begin{table}[htbp!]
\centering
\caption{Experimental results of SQWs on the example two-node graphs.}
\label{my-label}
\footnotesize
\begin{tabular}{l | l |l |l }
\hline
\textbf{$\{\alpha, \beta\}$}& \textbf{Initial state} & \textbf{Step number} &\textbf{Average fidelity}\\ 
\hline
$\{0.1, 0.9\}$ & $\ket{0}\ket{0}$ & 200 & 98.35$\pm$0.15\% \\
 & $\frac{1}{\sqrt{2}}\ket{0}(\ket{0}+\ket{1})$ & 200 & 99.38$\pm$0.03\% \\
 & $\frac{1}{\sqrt{2}}\ket{0}(\ket{0}+i\ket{1})$ & 200 & 99.13$\pm$0.04\% \\
\hline
$\{0.3, 0.7\}$ & $\ket{0}\ket{0}$ & 200 & 97.99$\pm$0.10\% \\
 & $\frac{1}{\sqrt{2}}\ket{0}(\ket{0}+\ket{1})$ & 200 & 99.31$\pm$0.04\% \\
 & $\frac{1}{\sqrt{2}}\ket{0}(\ket{0}+i\ket{1})$ & 200 & 98.71$\pm$0.05\% \\
\hline
$\{0.25, 0.25\}$ & $\ket{0}\ket{0}$ & 200 & 98.46$\pm$0.04\% \\
& $\frac{1}{\sqrt{2}}\ket{0}(\ket{0}+\ket{1})$ & 200 & 99.28$\pm$0.02\% \\
 & $\frac{1}{\sqrt{2}}\ket{0}(\ket{0}+i\ket{1})$ & 200 & 98.71$\pm$0.05\% \\
\hline
$\{0.5, 0.5\}$ & $\ket{0}\ket{0}$ & 200 & 98.48$\pm$0.04\%\\
 & $\frac{1}{\sqrt{2}}\ket{0}(\ket{0}+\ket{1})$ & 200 & 99.02$\pm$0.11\% \\
 & $\frac{1}{\sqrt{2}}\ket{0}(\ket{0}+i\ket{1})$ & 200 & 98.81$\pm$0.03\% \\
\hline
$\{0.43, 0.43\}$ & $\ket{0}\ket{0}$ & 200 & 98.02$\pm$0.04\% \\
 & $\frac{1}{\sqrt{2}}\ket{0}(\ket{0}+\ket{1})$ & 200 & 99.23$\pm$0.02\% \\
& $\frac{1}{\sqrt{2}}\ket{0}(\ket{0}+i\ket{1})$ & 200 & 98.88$\pm$0.03\% \\
\hline
$\{0.45, 0.45\}$ & $\ket{0}\ket{0}$ & 200 & 97.86$\pm$0.04\% \\
 & $\frac{1}{\sqrt{2}}\ket{0}(\ket{0}+\ket{1})$ & 200 & 99.12$\pm$0.04\% \\
 & $\frac{1}{\sqrt{2}}\ket{0}(\ket{0}+i\ket{1})$ & 200 & 98.75$\pm$0.03\% \\
\hline
$\{0.47, 0.47\}$ & $\ket{0}\ket{0}$ & 200 & 98.11$\pm$0.04\% \\
 & $\frac{1}{\sqrt{2}}\ket{0}(\ket{0}+\ket{1})$ & 200 & 99.30$\pm$0.02\% \\
 & $\frac{1}{\sqrt{2}}\ket{0}(\ket{0}+i\ket{1})$ & 200 & 98.84$\pm$0.03\% \\
\hline
\end{tabular}
\end{table}

\begin{figure}[htbp!] 
\centering    
\includegraphics[scale=0.8]{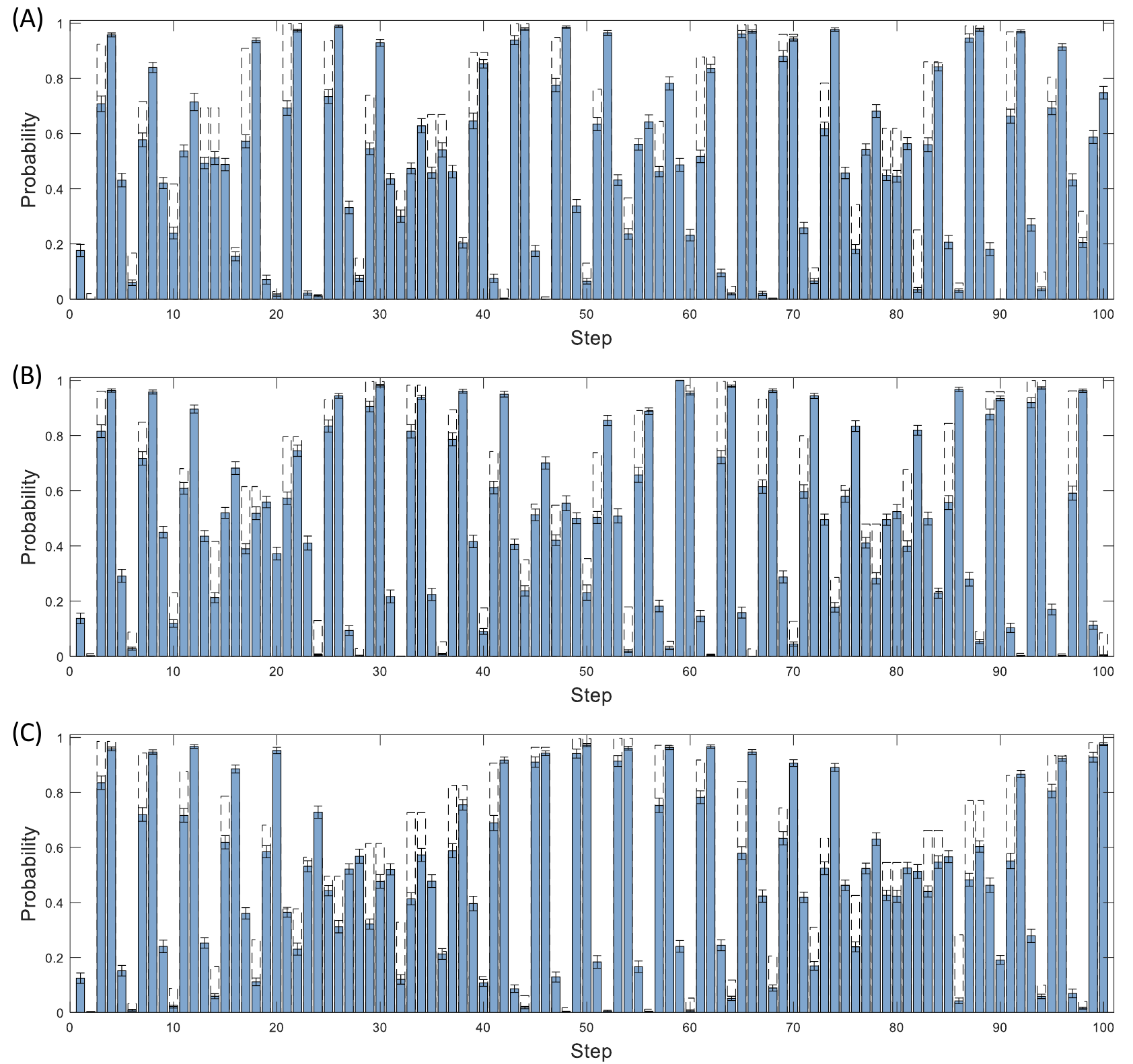}
\caption{Experimental obtained probability distributions of SQWs with the theoretical distribution overlaid. (A) Probability distribution (of the walker being at node 1) of the first 100 steps of SQW on the example two-node graph with $\alpha=\beta=0.43$ and an initial state of $\ket{00}$. (B) Probability distribution (of the walker being at node 1) of the first 100 steps of SQW on the example two-node graph with $\alpha=\beta=0.45$ and an initial state of $\ket{00}$. (C) Probability distribution (of the walker being at node 1) of the first 100 steps of SQW on the example two-node graph with $\alpha=\beta=0.47$ and an initial state of $\ket{00}$. }
\label{fig:SQWProbDis}
\end{figure}

\end{document}